%% file: root_old.tex
\documentclass[11pt,letterpaper]{article}

\usepackage[margin=1in]{geometry}
\usepackage{graphics} 
\usepackage{tikz}
\usetikzlibrary{positioning}
\usetikzlibrary{arrows}
\usepackage{natbib} 
\usepackage{epstopdf}
\usepackage{epsfig} 
\usepackage{subcaption}
\usepackage{psfrag}
\usepackage[width=.85\textwidth]{caption}
\usepackage{lipsum}
\usepackage{amsmath} 
\usepackage{amssymb}  
\usepackage{mathbbol}
\usepackage{cite}
\usepackage{mathrsfs}
\usepackage{url}
\usepackage{amsmath,amsfonts, amssymb,color}
\usepackage{amsthm}
\usepackage[width=.85\textwidth, font=footnotesize]{caption}
\usepackage[pdfpagelabels,pdfpagemode=None,breaklinks=true]{hyperref} \hypersetup{
     colorlinks   = true,
     citecolor    = blue
}
\renewcommand{\cite}{\citep}

\newtheorem{theorem}{Theorem}
\newtheorem{lemma}{Lemma}
\newtheorem{remark}{Remark}
\newtheorem{proposition}{Proposition}
\newtheorem{assumption}{Assumption}
\newtheorem{definition}{Definition}
\newtheorem{example}{Example}

\makeatletter
\renewcommand\bibsection%
{
  \section*{\refname
    \@mkboth{\MakeUppercase{\refname}}{\MakeUppercase{\refname}}}
}
\makeatother

\newcommand{\ignore}[1]{}

\DeclareMathOperator*{\argmin}{argmin}
\DeclareMathOperator*{\argmax}{argmax}
\DeclareMathOperator*{\avg}{avg}

\newcommand*{\TitleFont}{%
      \usefont{\encodingdefault}{\rmdefault}{}{n}%
      \fontsize{16}{20}%
      \selectfont}

\DeclareMathAlphabet{\mathcal}{OMS}{cmsy}{m}{n}      
      
\begin{document}

\title{\TitleFont Interdependent Security Games on Networks \\ under Behavioral Probability Weighting\thanks{A preliminary version of this work appeared in the proceedings of GameSec 2015 \cite{hota2015interdependent}.}}
\author{~Ashish~R.~Hota~and~Shreyas~Sundaram\thanks{The authors are with the School of Electrical and Computer Engineering, Purdue University. E-mail: $\{\text{ahota,sundara2}\}$@purdue.edu. This work is supported by a grant from the Purdue Research Foundation.}}%

\date{}
\maketitle

\begin{abstract}
We consider a class of interdependent security games on networks where each node chooses a personal level of security investment. The attack probability experienced by a node is a function of her own investment and the investment by her neighbors in the network. Most of the existing work in these settings considers players who are risk-neutral. In contrast, studies in behavioral decision theory have shown that individuals often deviate from risk-neutral behavior while making decisions under uncertainty. In particular, the true probabilities associated with uncertain outcomes are often transformed into perceived probabilities in a highly nonlinear fashion by the users, which then influence their decisions. In this paper, we investigate the effects of such behavioral probability weightings by the nodes on their optimal investment strategies and the resulting security risk profiles that arise at the Nash equilibria of interdependent network security games. We characterize graph topologies that achieve the largest and smallest worst case average attack probabilities at Nash equilibria in {\it Total Effort} games, and equilibrium investments in {\it Weakest Link} and {\it Best Shot} games.
\end{abstract}

\section{Introduction}
\label{section:introduction}
\input{introduction.tex} 

\section{Probability Weighting}
\label{section:probweighting}
\input{probweighting.tex} 

\section{Interdependent Security Games}
\label{section:securitygame}
\input{securitygame.tex} 

\section{Pure Nash Equilibria in Total Effort Games with Behavioral Probability Weighting}
\label{section:nashequilibriumexistence}
\input{heteroalpha.tex} 

\section{Effects of Network Structure}
\label{section:DegreeHeterogeneity}
\input{DegreeHeterogeneity.tex} 

\section{Comparative Statics in Weighting Functions}
\label{section:classesofgraphs}
\input{specificclasses.tex} 

\section{Weakest Link and Best Shot Games}
\label{section:weakestandbest}
\input{weakestandbest.tex} 

\section{Numerical Examples}
\label{section:numerical}
\input{numerical.tex} 

\section{Discussion and Conclusion}
We studied a class of interdependent security games on networks where the players exhibit certain behavioral attributes vis-a-vis their perception of attack probabilities, while making security investment decisions. We analyzed three canonical interdependent security game models, (i) {\it Total Effort}, (ii) {\it Weakest Link}, and (iii) {\it Best Shot} games. 

The Nash equilibria in {\it Total Effort} games have much richer structural properties under behavioral probability weighting than the corresponding equilibria with players that are true expectation maximizers. A sharper overweighting of small probabilities may disincentivize a node from reducing her investment when her neighbors make high investments. Underweighting of large probabilities leads to equilibria where the nodes are not completely unprotected, as opposed to equilibria without probability weighting. The effect of behavioral probability weighting is most beneficial (in terms of reducing the security risk) when the probability of successful attack is sufficiently high at the PNE. On the other hand, if the attack probability is only moderately high, players with weighting functions that are closer to linear have a more secure equilibrium. We obtained an upper bound on the expected fraction of nodes that are successfully attacked at any PNE, in terms of the average degree of the nodes. Furthermore, among the class of graphs with a given average degree $d_{\avg}$, the $d_{\avg}$-regular graph has a PNE which achieves this upper bound. Conversely, star graphs achieve the smallest average security risk upper bound among all connected graphs. 

When the graph is connected, all nodes make identical investments in a {\it Weakest Link} game, and there is a continuum of attack probabilities, independent of the graph structure, that can arise at a PNE. On the other hand, a strategy profile is a PNE in a {\it Best Shot} game if and only if the nodes making nonzero security investments form a maximal independent set. In both cases, equilibria are never completely unprotected under behavioral probability weighting.

\sloppy
\bibliographystyle{plainnat}
{\footnotesize \bibliography{refs}}

\input{appendix.tex} 

\end{document}

%% file: introduction.tex
Interdependent security games are a class of strategic games where the security risk faced by a player (often manifested as the probability of a successful attack) depends on her personal investment in security and the investments by other interacting players~\cite{laszka2014survey,kunreuther2003interdependent}. This is a broad framework to capture security interdependencies between independent stakeholders in (networked) cyber-physical systems. There is a large literature on this class of problems~\cite{laszka2014survey,kunreuther2003interdependent,varian2004system,manshaei2013game} motivated by applications in cybersecurity, airline security and epidemic risks. 

Much of the work in interdependent security games considers players who are risk-neutral, or are risk averse in the sense of classical expected utility theory~\cite{laszka2014survey}. On the other hand, there is a rich literature in decision theory and behavioral economics showing that human behavior consistently and significantly deviates from the predictions of classical expected utility theory~\cite{camerer2011advances}. While there have been some studies highlighting the significance of biases and irrationalities in human decision-making in information security domains~\cite{christin2011network,schneier2008psychology,garg2013heuristics}, theoretical analyses of deviations from classical notions of rational behavior are scarce in the literature on interdependent security games. Empirical investigations are also limited \cite{christin2012s}.

The goal of this paper is to initiate a rigorous investigation of the impact of behavioral decision-theoretic models in interdependent security games. In the context of security games, one of the most important behavioral deviations from the classical expected utility framework is the way individuals perceive the probability of an uncertain outcome (e.g., cyber attack).\footnote{There are also various behavioral characteristics that affect the perceived {\it values} of gains and losses~\cite{tversky1992advances,hota2014fragility}.} In particular, empirical studies show that individuals tend to overweight small probabilities and underweight large probabilities. Thus, the true probabilities are typically transformed in a highly nonlinear fashion into {\it perceived} probabilities, which are then used for decision-making~\cite{tversky1992advances,gonzalez1999shape}. These transformations are captured in the form of probability weighting functions.

In this paper, we analyze the effects of behavioral probability weighting on players' equilibrium strategies in interdependent security games on networks. We consider three canonical manifestations of the security risk in the forms of {\it Total Effort}, {\it Weakest Link} and {\it Best Shot} games. These game-theoretic models were first introduced in \cite{varian2004system} and have been studied extensively in the literature to model several scenarios in the cybersecurity domain, as described below.

In {\it Total Effort} games on networks, the probability of a successful attack on a node is an (affine) decreasing function of the average of the security investments by the node and her neighbors. The total effort externality has been studied as an abstraction of several cybersecurity problems \cite{grossklags2008security,grossklags2009uncertainty}. For instance, when an attacker tries to slow down file transfers in peer-to-peer networks, the success of the attack depends on the aggregate effort of all the participating agents \cite{grossklags2008security}. Similar externalities arise when underinvestment in security by a user potentially causes increasing spam activity for others who communicate with her \cite{laszka2014survey}. In~\cite{miura2008security,nguyen2009stochastic}, the authors consider a similar formulation, where the security risk faced by a node is a weighted linear combination of her own investment and the investments by her neighbors. The authors discuss multiple settings where such externalities arise, such as in web authentication and spam verification. Amin et. al~\cite{amin2013security} study a related setting, where a set of independent control systems interact over a shared communication network, and the failure probability is a function of the number of controllers who have invested in security.\footnote{Security externalities similar to the Total Effort formulation have also been studied in the broader interdependent security game literature in the context of inefficiency of equilibria~\cite{jiang2011bad}, incomplete information about the network topology~\cite{pal2011modeling} and cyber insurance~\cite{schwartz2013cyber}. In all these settings, the security risk faced by a node is determined by the actions of the node and her immediate neighbors. This is different from the line of work that models epidemic risks and cascading failures spreading over the network~\cite{nowzari2016analysis}.} 

\begin{table*}[t]
\captionsetup{width=0.8\textwidth}
\centering
\begin{center}
  \begin{tabular}{ | c  | c | c | }
    \hline
    Externality &  Impact of Network Structure & Impact of Weighting Function  \\ \hline
     & The expected fraction of nodes that are & In degree-regular graphs, the  \\ 
    Total  & successfully attacked at a PNE is highest & interior PNE is more secure under   \\ 
    Effort & in degree-regular graphs, among & probability weighting when the \\ 
    & all graphs with a given average degree. & graph is sufficiently dense. \\ \hline
     & In a connected graph, all & The attack probabilities that  \\ 
    Weakest & nodes have identical & arise at a PNE have values  \\ 
   Link & investments at a PNE. & close to $1$ and/or $0$, depending \\ & & on the game parameters.
   \\ \hline
     & Nodes with nonzero investments & Under probability weighting, \\
    Best & at a PNE form a & the PNE is never fully insecure, \\ 
   Shot & maximal independent set. & i.e., there always exist node(s) \\
 & & with nonzero investment(s).   \\ \hline
  \end{tabular}
\caption[width=.8\linewidth]{Summary of main results of the pure Nash equilibrium (PNE) characteristics for different attack probability functions.\label{table:summary}}
\end{center}
\end{table*}

In the {\it Weakest Link} game, a node is only as secure as the least secure node in her neighborhood, while in the {\it Best Shot} game, the player with the maximum investment in the neighborhood must be successfully attacked for the attack on a node to be successful. Weakest link externalities are prevalent in cybersecurity domains; successful breach of one subsystem often increases the vulnerability of connected subsystems by giving the attacker increased access to otherwise restricted parts. Best shot externalities arise in cyber-physical systems that have built-in redundancies. The attacker must breach the most secure subsystem for the attack to be successful. Best shot externalities also arise in censorship resilient communication, where information is available to a node as long as one of her neighbors possesses that information \cite{johnson2010are}. We summarize our main findings on the effects of network structure and probability weighting on players' strategies and equilibrium attack probabilities for all three games in Table~\ref{table:summary}.

Security investments often exhibit characteristics of a public good \cite{varian2004system}, and the different risk externalities described above have indeed been studied in the context of public good games \cite{hirshleifer1983weakest}. There is also a growing interest in networked public goods in recent years \cite{galeotti2010network,bramoulle2014strategic}. Our analysis of behavioral probability weighting complements this line of research. 

%% file: probweighting.tex
\begin{figure}[t]
	\centerline{\includegraphics[width=7cm,height=4.5cm]{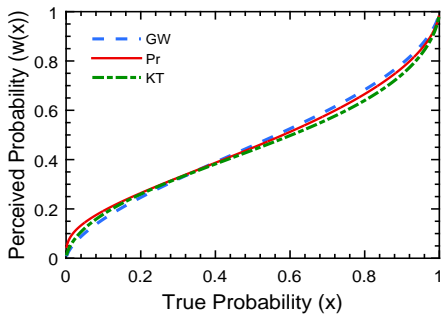}}
	\caption{Shape of the probability weighting function for different parametric forms proposed by Gonzalez and Wu (GW)~\cite{gonzalez1999shape}, Prelec (Pr)~\cite{prelec1998probability} and Kahneman and Tversky (KT)~\cite{tversky1992advances}.}
	\label{fig:allthree}
\end{figure}

As discussed in the previous section, our focus in this paper will be on understanding the effects of nonlinear weighting of true probabilities by individuals while making decisions under risk. Such weightings have been comprehensively studied in the behavioral economics and psychology literature~\cite{camerer2011advances}, and more recently in wireless communications \cite{li2014users} and the smart grid \cite{saad2016toward}. Behavioral perception of probabilities by human decision makers have certain fundamental characteristics, including (i) {\it possibility effect}: overweighting of probabilities very close to $0$, (ii) {\it certainty effect}: underweighting of probabilities very close to $1$, and (iii) {\it diminishing sensitivity} from the end points $0$ and $1$. These characteristics are usually captured by an inverse S-shaped weighting function $w :[0,1] \to [0,1]$; prominent parametric forms of such weighting functions were proposed by Kahneman and Tversky~\cite{tversky1992advances}, Gonzalez and Wu~\cite{gonzalez1999shape}, and Prelec~\cite{prelec1998probability}, and are illustrated in Figure~\ref{fig:allthree}.     

For our general analysis of the Nash equilibrium in Total Effort games in Section~\ref{section:nashequilibriumexistence} and Weakest Link and Best Shot games in Section~\ref{section:weakestandbest}, we will assume that the derivative of the weighting function satisfies the following properties.

\begin{assumption}\label{assumption:weightingfunction}
The probability weighting function $w(x)$ has the following properties. 
\begin{enumerate}
\item[1.] $w'(x)$ has a unique minimum for $x \in (0,1)$ denoted as $\mathbf{x}_{\min,w} := \argmin_{x \in [0,1]} w'(x)$, and $w''(\mathbf{x}_{\min,w}) = 0$.
\item[2.] $w(x)$ is strictly concave for $x \in [0,\mathbf{x}_{\min,w})$, and is strictly convex for $x \in (\mathbf{x}_{\min,w},1]$.
\item[3.] $w'(\epsilon) \to \infty$ as $\epsilon \to 0$, and $w'(1-\epsilon) \to \infty$ as $\epsilon \to 0$.
\end{enumerate}
\end{assumption}

We impose the following additional requirements on the weighting functions in order to obtain certain results on the effects of network structure on the equilibria in Section~\ref{section:DegreeHeterogeneity}.

\begin{assumption}\label{assumption:degreeheterogeneity} 
The probability weighting function $w(x)$ has the following properties.
\begin{enumerate}
\item[1.] $\frac{w''(x)}{w'(x)} < \frac{1}{1-x}$ for $x \in (\mathbf{x}_{\min,w},1)$.
\item[2.] $w'(x)$ is strictly convex for $x \in (\mathbf{x}_{\min,w},1)$.
\end{enumerate}
\end{assumption}  

\begin{remark}
Assumption~\ref{assumption:weightingfunction} and~\ref{assumption:degreeheterogeneity} hold true for the parametric forms of the weighting functions proposed by Kahneman and Tversky~\cite{tversky1992advances}, Gonzalez and Wu~\cite{gonzalez1999shape}, and Prelec~\cite{prelec1998probability} for the ranges of parameter values under which these functions have an inverse-S shape (Figure~\ref{fig:allthree}). In particular, the assumptions hold for the ranges of parameter values estimated from empirical studies on human subjects; \cite{booij2010parametric} contains a review of several such studies.
\end{remark}

For our results on the effect of the intensity of overweighting and underweighting of the weighting functions on the equilibrium attack probabilities in Section~\ref{section:classesofgraphs}, we will need to consider a specific parametric form of the weighting function. For that purpose, we use the single-parameter Prelec weighting function from~\cite{prelec1998probability} due to its analytical tractability. In particular, if the true probability of an outcome is $x$, the Prelec weighting function is given by
\begin{equation}
w(x) = \exp(-(-\ln(x))^\alpha), \text{\qquad} x \in [0,1],
\label{eq:prelec}
\end{equation}
where $\exp(\cdot)$ is the exponential function. The parameter $\alpha \in (0,1)$ controls the curvature of the weighting function. For $\alpha=1$, we have $w(x)=x$, i.e., the weighting function is linear. For smaller $\alpha$, the function $w(x)$ has a sharper overweighting of low probabilities and underweighting of high probabilities. A useful property of this function is that regardless of the value of $\alpha$, $\mathbf{x}_{\min,w} = \frac{1}{e}$, and $w(\frac{1}{e}) = \frac{1}{e}$. In other words, $w''(x) = 0$ at $x = \frac{1}{e}$. The minimum value of $w'(x)$ is $w'(\frac{1}{e}) = \alpha$. 

%% file: securitygame.tex
In this paper, we consider interdependent security games on networks. Let $\mathcal{G} = \{\mathcal{V},\mathcal{E}\}$ denote an undirected network (or graph) with $\mathcal{V}$ being the set of nodes with $|\mathcal{V}| = n$. Each node is an independent decision-maker (player of the game) representing, for instance, an entity in a cyber-physical system. The security investment by node $i$ is denoted as $s_i$, with $s_i \in [0,1]$. The security risk or attack probability experienced by a node is a function of her investment $s_i$, and the investment by her direct neighbors. We denote the set of neighbors of node $i$ as $\mathcal{N}(i)$, and the investment profile of all nodes in $\mathcal{N}(i)$ as the vector $\mathbf{s}_{\mathcal{N}(i)}$. The true probability of a successful attack on node $i$ is given by $f_i(s_i, \mathbf{s}_{\mathcal{N}(i)}) \in [0,1]$, for some function $f_i$. Node $i$ incurs a cost-per-unit of security investment of $c_i \in \mathbb{R}_{\ge 0}$, and if the attack is successful, she incurs a loss of $L_i \in \mathbb{R}_{> 0}$. Her expected utility (under the true probability of successful attack) is then 
\begin{equation}\label{eq:riskneutralutility}
\mathbb{E}u_i (s_i,\mathbf{s}_{\mathcal{N}(i)}) = -L_i f_i(s_i,\mathbf{s}_{\mathcal{N}(i)}) - c_i s_i.
\end{equation} 
For ease of notation, we define the {\it extended neighborhood} of node $i$, denoted $\bar{\mathcal{N}}(i)$, as the set of nodes including herself and her neighboring nodes, i.e., $\bar{\mathcal{N}}(i) \triangleq \mathcal{N}(i) \cup \{i\}$. We denote the size of the extended neighborhood of a node $i$ as,
\begin{equation}\label{equation:di}
d_i \triangleq 1+|\mathcal{N}(i)|.
\end{equation}

In this work, we consider three canonical models of interdependent security games initially presented in \cite{grossklags2008security}. The models differ in the attack probability function $f_i(s_i,\mathbf{s}_{\mathcal{N}(i)})$ as described below. 
\begin{itemize}
\item {\it Total Effort}: $f_i(s_i,\mathbf{s}_{\mathcal{N}(i)}) = 1-\frac{1}{d_i}(\sum_{j \in \bar{\mathcal{N}}(i)} s_j)$.
\item {\it Weakest Link}:  $f_i(s_i,\mathbf{s}_{\mathcal{N}(i)}) = 1-\min_{j \in \bar{\mathcal{N}}(i)} s_j$.
\item {\it Best Shot}:  $f_i(s_i,\mathbf{s}_{\mathcal{N}(i)}) = 1-\max_{j \in \bar{\mathcal{N}}(i)} s_j$.
\end{itemize}
Note that in prior works (e.g. \cite{grossklags2008security}), the games were defined over complete graphs, i.e., $\bar{\mathcal{N}}(i) = \mathcal{V}$, while we consider more general graph topologies. Most of our analysis will focus on the {\it Total Effort} attack probability function, since the results also have potential implications for other classes of security games considered in the literature~\cite{miura2008security,nguyen2009stochastic}. Since the focus of the present work is to understand the effects of behavioral probability weighting functions and node degrees on the Nash equilibrium security levels, we only focus on the case where the security risk of a node is influenced in an identical way by all of her neighbors' investments.

We formally define the notion of an equilibrium and {\it best response} in interdependent security games as follows.
\begin{definition}\label{def:pne}
A strategy profile $\{s^*_1,s^*_2,\ldots,s^*_n\}$ is a pure Nash equilibrium (PNE) if for every player $i \in \{1,2,\ldots,n\}$, $\mathbb{E}u_i (s'_i,\mathbf{s}^*_{\mathcal{N}(i)}) \leq \mathbb{E}u_i (s^*_i,\mathbf{s}^*_{\mathcal{N}(i)}), \forall s'_i \in [0,1]$.
\end{definition}

\begin{definition}\label{def:bestresponse}
The {\it best response} of player $i$ at a given investment profile $\mathbf{s}_{\mathcal{N}(i)}$ by her neighbors is the set $b_i(\mathbf{s}_{\mathcal{N}(i)}) \triangleq \argmax_{s_i \in [0,1]}  \mathbb{E}u_i(s_i,\mathbf{s}_{\mathcal{N}(i)})$. A strategy profile $\{s^*_1,s^*_2,\ldots,s^*_n\}$ is a pure Nash equilibrium if and only if $s^*_i \in b_i(\mathbf{s}^*_{\mathcal{N}(i)})$ for every player $i \in \{1,2,\ldots,n\}$.
\end{definition}

In other words, a PNE exists if the vector of best response mappings $[b_1(\cdot),b_2(\cdot),\ldots,b_n(\cdot)]$ possesses a fixed point \cite{osborne1994course}.
 
\subsection{Equilibria without probability weighting}

To establish a baseline, we present the following proposition that describes the main results from~\cite{grossklags2008security} regarding the properties of the best response of a player with $w(x)=x$ in a {\it Total Effort} game; while that paper only considered complete graphs, the result extends directly to general graphs. We refer to a player with $w(x)=x$ as a {\it true expectation maximizer} as she maximizes her expected utility~\eqref{eq:riskneutralutility} without any behavioral probability weighing.  

\begin{proposition}\label{proposition:riskneutralresult}
Consider a player $i$ with $w_i(x) = x$ in a {\it Total Effort} game with extended neighborhood size $d_i$. Then, her best response is $s^*_i = 1$ when $\frac{d_ic_i}{L_i} < 1$, and $s^*_i = 0$ if $\frac{d_ic_i}{L_i} > 1$. In the special case where $\frac{d_ic_i}{L_i} = 1$, any investment $s_i \in [0,1]$ is an optimal strategy.
\end{proposition}

Note that, except for the pathological case where $\frac{d_ic_i}{L_i} = 1$, the best response of a player is to either fully protect herself or remain completely unprotected. As the size of her neighborhood increases, her best response jumps from investing $1$ to investing $0$. Such behavior arises since the marginal utility of a true expectation maximizer in~\eqref{eq:riskneutralutility} is independent of her strategy and the strategies of the players in her neighborhood. 

When the cost parameters $c_i$ and $L_i$ are homogeneous across players, only a set of nodes with small enough degrees make an investment of $1$, while the high degree nodes invest zero in any Nash equilibrium of the {\it Total Effort} game. Furthermore, in degree regular graphs, the only equilibrium that arises has all players investing $0$ or investing $1$, i.e., all players are either fully secure or fully unprotected.  

In our analysis, we will show that under behavioral probability weighting, both the best responses and the equilibria have much richer structural properties and vary more smoothly with the weighting parameters and the network structure.

%% file: heteroalpha.tex
In this section, we consider {\it Total Effort} games on networks with player-specific probability weighting functions $w_i(\cdot)$ (satisfying Assumption~\ref{assumption:weightingfunction}) and cost parameters $c_i$ and $L_i$. We first prove the existence of a PNE in this class of games by establishing the existence of a fixed point of the best response mapping (see Definition~\ref{def:bestresponse}).

With probability weighting, the expected utility of player $i$ with investment $s_i \in [0,1]$ is given by
\begin{equation}\label{eq:weightedutility}
\mathbb{E}u_i(s_i,\mathbf{s}_{\mathcal{N}(i)}) = -L_i w_i\left(1-\frac{s_i + \bar{s}_{-i}}{d_i}\right)-c_i s_i,
\end{equation}
where $\bar{s}_{-i} = \sum_{j \in \mathcal{N}(i)} s_j$ is the total investment in security by the neighbors of $i$. The quantity $d_i$ is the size of the extended neighborhood of player $i$ as defined in~\eqref{equation:di}. The marginal utility is given by
\begin{equation}
\frac{\partial \mathbb{E}u_i}{\partial s_i} = \frac{L_i}{d_i}w'_i\left(1-\frac{s_i + \bar{s}_{-i}}{d_i}\right)-c_i.
\label{eq:first_order}
\end{equation}
The solutions of $\frac{\partial \mathbb{E}u_i}{\partial s_i} =  0$ (if any) satisfy the first order necessary condition of optimality, and are therefore candidate solutions for player $i$'s best response. Note that $\left(1-\frac{s_i + \bar{s}_{-i}}{d_i}\right)$ is the true attack probability faced by player $i$ (without probability weighting). 

\begin{figure}[t]
	\centerline{\includegraphics[width=7cm,height=4.5cm]{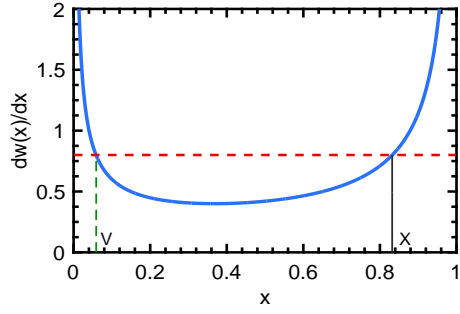}}
	\caption{Interior solutions of $w'(x) = \frac{dc}{L}$ are denoted by $V$ and $X$. In this example, $\frac{dc}{L} = 0.8$ and is shown by the horizontal line.  The Prelec weighting parameter is $\alpha = 0.4$.}
	\label{fig:interiorfoc}
\end{figure}

We illustrate the nature of solutions of the first order condition~\eqref{eq:first_order} in Figure~\ref{fig:interiorfoc} for a Prelec weighting function with parameter $\alpha=0.4$. One can see from the figure and \eqref{eq:first_order} that if $\frac{d_ic_i}{L_i} \leq w'_i(\mathbf{x}_{\min,w_i}) =  \min_{x \in [0,1]} w'_i(x)$, we have $\frac{\partial \mathbb{E}u_i}{\partial s_i} \geq 0$ for $s_i \in [0,1]$, and investing $1$ is the only best response of a player irrespective of the strategies of her neighbors.\footnote{When $\frac{d_ic_i}{L_i} = w'_i(\mathbf{x}_{\min,w_i})$, there is a unique solution to $\frac{\partial \mathbb{E}u_i}{\partial s_i} = 0$ at $\mathbf{x}_{\min,w_i}$, and $\frac{\partial \mathbb{E}u_i}{\partial s_i} > 0$ for all other $s_i$. The player prefers to invest $1$ in this case. The proof is similar to Case 4 in the proof of Lemma~\ref{lemma:bestresponse}.} 

Now suppose $\frac{d_ic_i}{L_i} > w'_i(\mathbf{x}_{\min,w_i})$. In this case, the first order condition $w'_i(x) = \frac{d_ic_i}{L_i}$ has two distinct interior solutions corresponding to true attack probabilities $V_i < \mathbf{x}_{\min,w}$ and $X_i > \mathbf{x}_{\min,w}$, as illustrated in Figure~\ref{fig:interiorfoc}. Note that as the degree of the node increases, $X_i - V_i$ increases as well.

When $\bar{s}_{-i}$ is the total investment by the neighbors of player $i$, player $i$'s strategy can change her true attack probability in the interval 
\begin{equation}\label{eq:bestresponseinterval}
\mathcal{X}(\bar{s}_{-i}) \triangleq \left[1-\frac{1 + \bar{s}_{-i}}{d_i},1-\frac{ \bar{s}_{-i}}{d_i}\right].
\end{equation}
In other words, when the extended neighborhood size is $d_i$, each player can directly change the probability of successful attack by at most $\frac{1}{d_i}$. We will make the following assumptions.  

\begin{assumption}\label{assumption:largeN}
For any player $i$ with $\frac{d_ic_i}{L_i} > w'_i(\mathbf{x}_{\min,w_i})$, let the size of the extended neighborhood ($d_i$), and the cost parameters ($c_i$ and $L_i$) be such that, (1) $X_i - V_i  > \frac{1}{d_i}$, (2) $V_i < \frac{1}{d_i}$ and (3) $w_i(\frac{1}{d_i}) < \frac{c_i}{L_i}$.
\end{assumption}

The first condition implies that at a given $\bar{s}_{-i}$, player $i$ does not contain both $V_i$ and $X_i$ in $\mathcal{X}(\bar{s}_{-i})$. This is required for maintaining continuity of the best response as $\bar{s}_{-i}$ varies. We will discuss the implications of the last two assumptions later in the analysis. Unless otherwise stated, the results of this section hold under Assumption~\ref{assumption:weightingfunction} for the probability weighting functions and Assumption~\ref{assumption:largeN} for the neighborhood sizes and the cost parameters.

We start with the following characterization of the best response of a player. In particular, we show that under Assumptions~\ref{assumption:weightingfunction} and \ref{assumption:largeN}, the best response of a player is unique, continuous, and monotonically decreasing in the aggregate investment by other nodes in her neighborhood. The proof is presented in Appendix~\ref{appendix:existence}.

\begin{lemma}\label{lemma:bestresponse}
Suppose $\frac{d_i c_i}{L_i} > \min_{x \in [0,1]} w'_i(x)$ for a player $i$. Then, for a given aggregate investment $\bar{s}_{-i}$ by the neighboring nodes, the best response of player $i$ is given by,
\begin{equation*}
b_i(\bar{s}_{-i}) =
\begin{cases}
1 & \text{when } \bar{s}_{-i} \leq d_i(1-X_i) - 1 \\
0 & \text{when } \bar{s}_{-i} \geq d_i(1-X_i), \\
d_i(1-X_i) - \bar{s}_{-i} & \text{otherwise}.
\end{cases}
\end{equation*}
where $X_i$ is the solution to $w'_i(x) = \frac{d_ic_i}{L_i}$ for $x \in (\mathbf{x}_{\min,w_i},1]$.
\end{lemma}

We use the properties of the best response proven in the above lemma to establish the existence of PNE in {\it Total Effort} games on networks under behavioral probability weighting. 

\begin{theorem}\label{theorem:existencemain}
Consider a {\it Total Effort} game on a graph where the weighting functions of the players satisfy Assumption~\ref{assumption:weightingfunction} and the cost parameters and neighborhood sizes satisfy Assumption~\ref{assumption:largeN}. Then this game admits a pure Nash equilibrium.
\end{theorem}
\begin{proof}
For a player $i$, if $\frac{d_ic_i}{L_i} \leq w'_i(\mathbf{x}_{\min,w_i})$, her best response is to invest $1$ regardless of the investments by her neighbors. Otherwise, the best response is unique and continuous in the strategies of her neighbors (from Lemma~\ref{lemma:bestresponse}). In addition, the strategy space of each player is $[0,1]$, which is compact and convex. Thus, according to Brouwer's fixed point theorem \cite{ok2007real}, there exists a fixed point of the best response mapping, which corresponds to a PNE.
\end{proof}

As a consequence of Lemma~\ref{lemma:bestresponse}, at a PNE, the investments in the extended neighborhood of any player $i$ for whom $\frac{d_ic_i}{L_i} > w'(\mathbf{x}_{\min,w_i})$ can be expressed as,
\begin{equation}\label{eq:projection}
s^*_i = \min ( \max (d_i(1-X_i)-\bar{s}^*_{-i},0),1).
\end{equation}
In particular, we have
\begin{equation}\label{eq:neighborhoodaggregate}
\begin{split}
1 + \bar{s}^*_{-i} < d_i(1-X_i) \implies & s^*_i = 1
\\ s^*_i + \bar{s}^*_{-i} = d_i(1-X_i) \implies & s^*_i \in [0,1]
\\ \bar{s}^*_{-i} > d_i(1-X_i) \implies & s^*_i = 0,
\end{split}
\end{equation}
where $w'_i(X_i) = \frac{d_ic_i}{L_i}$ with $X_i \in (\mathbf{x}_{\min,w},1)$. The converse of the second identity above holds when $s^*_i \in (0,1)$. 

\begin{remark}[Equilibrium computation]
Recently, the authors in \cite{gharesifard2015convergence} showed that when the best responses of the players are given by Lemma~\ref{lemma:bestresponse}, continuous best response dynamics converge to a PNE strategy profile. Furthermore, a strategy profile that satisfies \eqref{eq:neighborhoodaggregate} can be computed by solving a Linear Complementarity Program; a similar result was obtained in \cite{miura2008security}. We present an expanded discussion in Appendix~\ref{appendix:lcp}.
\end{remark}

In general, strategy profiles that satisfy equation \eqref{eq:projection} need not be unique, and therefore, we need not have a unique PNE. However, in the special case of complete graphs (the classical setting for {\it Total Effort} games~\cite{grossklags2008security}), we show that the strategy profiles at the Nash equilibria are unique up to the true equilibrium attack probability experienced by the players. 

\begin{proposition}\label{proposition:totaleffortNEheterounique}
Consider a {\it Total Effort} game on a complete graph where each player $i$ has a player-specific weighting function $w_i(\cdot)$ satisfying Assumption~\ref{assumption:weightingfunction}, and cost ratio $\frac{c_i}{L_i}$ satisfying Assumption~\ref{assumption:largeN}. Then all Nash equilibria have the same (true) probability of successful attack at the nodes. 
\end{proposition}

The proof is presented in Appendix~\ref{appendix:existence}.

If the strategy profile at each neighborhood satisfies the second identity of \eqref{eq:neighborhoodaggregate}, then we refer to such a PNE as an {\it interior equilibrium}. In other words, at an interior equilibrium, the strategy profile $\mathbf{s}^*$ satisfies, 
\begin{equation}
(A+I_n)\mathbf{s}^* = \mathbf{d} \circ (\mathbf{1}-\mathbf{X}),
\label{eq:interiorequilibrim}
\end{equation} 
where $A$ is the adjacency matrix of the graph, $\mathbf{d}$ is the vector of neighborhood sizes, $\mathbf{1}$ is the all-ones vector, $\mathbf{X}$ is the vector of $X_i$'s of the players, and ``$\circ$" denotes the Hadamard (element-wise) product. Thus, at every interior equilibrium, the true attack probability faced by each node $i$ is $X_i$. The existence of an interior equilibrium is not always guaranteed, except in certain special cases, such as when players are homogeneous and all nodes in the graph have identical degrees (i.e., the graph is {\it degree-regular}). 

\begin{proposition}\label{proposition:degreeregular}
Consider a degree-regular graph with degree $d-1$ and homogeneous players such that $\frac{dc}{L} > w'(\mathbf{x}_{\min,w})$. Then the symmetric strategy profile where each node invests $1-X$ constitutes an interior PNE, where $w'(X) = \frac{dc}{L}, X \in (\mathbf{x}_{\min,w},1]$. 
\end{proposition}

The proof is straightforward by substituting $s^* = 1-X$ for every player in equation~\eqref{eq:neighborhoodaggregate}.

\subsection{Existence of secure equilibrium}

Assumption~\ref{assumption:largeN} ensures that the best response of a player remains continuous in the strategies of other players, which helps us establish the existence of PNE in Theorem~\ref{theorem:existencemain}. It is possible to also show the existence of a PNE where all players invest $1$ at equilibrium, when the second and third conditions of Assumption~\ref{assumption:largeN} do not hold for all the players. 
 
\begin{proposition}
Suppose that for every player $i$ in a {\it Total Effort} game, either $V_i \geq \frac{1}{d_i}$ or $w_i(\frac{1}{d_i}) > \frac{c_i}{L_i}$. Then there exists a PNE where all players invest $1$.  
\end{proposition}

\begin{proof}
Consider a player $i$ with $V_i \geq \frac{1}{d_i}$, and assume that all of her neighboring nodes are investing $1$. Then, the marginal utility $\frac{\partial \mathbb{E}u_i}{\partial s_i}$ of player $i$ is nonnegative over the interval $\mathcal{X} = \left[0,\frac{1}{d_i}\right]$, and her best response is to invest $1$. 

Similarly, when $V_i < \frac{1}{d_i}$, and $w_i(\frac{1}{d_i}) > \frac{c_i}{L_i}$ for player $i$, then her optimal investment is $1$ when $\bar{s}_{-i} = d_i - 1$. This follows from Case 4 of in the proof of Lemma~\ref{lemma:bestresponse} as substituting $\bar{s}_i = d_i-1$ yields $\mathbb{E}u_i(1,\mathbf{s}_{\mathcal{N}(i)}) - \mathbb{E}u_i(0,\mathbf{s}_{\mathcal{N}(i)}) > 0$ when $w_i(\frac{1}{d_i}) > \frac{c_i}{L_i}$. 
\end{proof}

Both $V_i \geq \frac{1}{d_i}$ and $w_i(\frac{1}{d_i}) > \frac{c_i}{L_i}$ occur when the weighting function sufficiently overweights very small attack probabilities. Overweighting of small attack probabilities discourages a player from reducing her investment, even when all other players are fully secure, as it would lead to a large {\it perceived} increase in attack probabilities from a relatively secure state. This results in a fully secure equilibrium. 

While such a fully secure PNE possibly coexists with other equilibria, identifying conditions under which such a PNE exists has potential implications for designing incentive mechanisms to encourage users to achieve a secure PNE. 

%% file: DegreeHeterogeneity.tex
In this section, we focus on understanding the effect of network structure, vis-a-vis the degrees of the nodes, on the security investments and attack probabilities at the PNE of {\it Total Effort} games. We consider players with homogeneous weighting functions and cost parameters in order to isolate the effects of their degrees on their investments. We will use the characterization of the PNE strategy profile given in equation~\eqref{eq:neighborhoodaggregate}. For the analysis in this section, we assume that the weighting functions satisfy both Assumptions~\ref{assumption:weightingfunction} and \ref{assumption:degreeheterogeneity}.

When the weighting functions and the cost parameters are homogeneous across the players, then the quantity $d_i (1-X_i)$ is only a function of the size of the extended neighborhood $d_i$. Recall that $w'(X_i) = \frac{d_ic}{L}$, and $X_i > \mathbf{x}_{\min,w}$. The properties of $X_i$ as a function of $d_i$ are the basis of the analytical results of this section. 

\begin{remark}\label{remark:increasingX}
Note that $X_i$ is an increasing function of $d_i$, as $w'(x)$ is strictly increasing in $x$ for $x \in (\mathbf{x}_{\min,w},1]$, as illustrated in Figure~\ref{fig:interiorfoc}.
\end{remark}

\subsection{Investments by nodes with overlapping neighborhoods}

Without probability weighting, Proposition~\ref{proposition:riskneutralresult} indicated that a lower degree node always invests at least as much as a higher degree node. This monotonicity does not hold in general under behavioral probability weighting. Nonetheless, we can prove certain monotonicity properties when $d(1-X)$ is a monotonically decreasing function of $d$, which holds under Assumption~\ref{assumption:degreeheterogeneity} as shown below.

\begin{lemma}\label{lemma:doneminusx}
Let $\frac{c}{L} > \min_{x \in [0,1]} w'(x)$ so that the quantity $X$ is defined for $d \geq 1$. If the probability weighting function satisfies $\frac{w''(x)}{w'(x)} < \frac{1}{1-x}$ for $x \in (\mathbf{x}_{\min,w},1)$, then $d(1-X)$ is monotonically decreasing in $d$.
\end{lemma}

The proof of Lemma~\ref{lemma:doneminusx} is presented in Appendix~\ref{section:appendixDegreeHeterogeneity}. We now prove the following result using Lemma~\ref{lemma:doneminusx}. Recall that $\bar{\mathcal{N}}(i)$ denotes the extended neighborhood of node $i$.  

\begin{proposition}\label{proposition:extended_neighborhood}
Consider a {\it Total Effort} game on a network with homogeneous players whose weighting functions satisfy Assumptions~\ref{assumption:weightingfunction} and \ref{assumption:degreeheterogeneity}. Let the cost parameters satisfy Assumption~\ref{assumption:largeN}, and let $\frac{c}{L} > \min_{x \in [0,1]} w'(x)$. If $\bar{\mathcal{N}}(i) \subset \bar{\mathcal{N}}(j)$, then at any PNE, $s^*_i \geq s^*_j$.
\end{proposition}

\begin{proof}
Consider any equilibrium strategy profile where player $j$ invests $s^*_j > 0$, for otherwise, the result holds trivially. Then from equation~\eqref{eq:neighborhoodaggregate}, and from the fact that $\bar{\mathcal{N}}(i) \subset \bar{\mathcal{N}}(j)$, we obtain 
\begin{equation*}
\sum_{k \in \bar{\mathcal{N}}(i)} s^*_k \leq \sum_{k \in \bar{\mathcal{N}}(j)} s^*_k \leq d_j(1-X_j) < d_i(1-X_i),
\end{equation*}
which follows from the monotonicity of $d(1-X)$ shown in Lemma~\ref{lemma:doneminusx}. As a result, from~\eqref{eq:neighborhoodaggregate}, we must have $s^*_i = 1$.   
\end{proof}

From Lemma~\ref{lemma:doneminusx}, we know that $d(1-X)$ is a decreasing function of $d$. From the PNE characterization in~\eqref{eq:neighborhoodaggregate}, node $i$ invests such that the aggregate investment in her extended neighborhood is $d_i(1-X_i)$ subject to her investment being within $[0,1]$. For a node with a few neighbors, the desired investment $d(1-X)$ is larger, and therefore, the node has to increase her personal investment. On the other hand, for a node with large degree, $d(1-X)$ is smaller and she has a large number of neighbors to rely on to meet the target investment.

\subsection{Upper bound on average probability of successful attack at equilibrium}

For a PNE strategy profile $\mathbf{s}^*$, we denote the average (true) probability of successful attack as
\begin{equation}
\Phi(\mathbf{s}^*) := \frac{1}{n} \sum^n_{i=1} \left(1-\frac{\sum_{j \in \bar{\mathcal{N}}(j)} s^*_j}{d_i}\right).
\end{equation}
Note that $\Phi(\mathbf{s}^*)$ is also equal to the expected fraction of nodes that are successfully attacked under the PNE strategy profile $\mathbf{s}^*$. We obtain an upper bound on $\Phi(\mathbf{s}^*)$ when $1 - X_i < \frac{1}{d_i}$ for every player $i$, with the proof presented in Appendix \ref{section:appendixDegreeHeterogeneity}.

\begin{proposition}\label{proposition:upperbound}
Consider a {\it Total Effort} game on a graph with homogeneous players who satisfy Assumptions~\ref{assumption:weightingfunction}, \ref{assumption:degreeheterogeneity} and \ref{assumption:largeN}. In addition, suppose $\frac{d_ic}{L} > \min_{x \in [0,1]} w'(x)$ and $1 - X_i < \frac{1}{d_i}$ for every player $i$. Then at any PNE with strategy profile $\mathbf{s}^*$,
\begin{enumerate}
\item the attack probability at node $i$ is at most $X_i$, and 
\item $\Phi(\mathbf{s}^*) \leq \frac{1}{n}\sum^n_{i=1} X_i$. 
\end{enumerate}
Furthermore, if there exists an interior PNE with strategy profile $\mathbf{s}^*_{\mathcal{I}}$, then $\Phi(\mathbf{s}^*_{\mathcal{I}}) = \frac{1}{n}\sum^n_{i=1} X_i$.
\end{proposition} 

\begin{remark}
If there is a leaf node $i$ with $d_i=2$ that satisfies Assumption~\ref{assumption:largeN}, then we must have $1-X_i < \frac{1}{2}$; otherwise $X_i - V_i < \frac{1}{2}$, which violates the first condition of Assumption~\ref{assumption:largeN}. Since $d(1-X)$ is decreasing in $d$ (according to Lemma~\ref{lemma:doneminusx} under Assumption~\ref{assumption:degreeheterogeneity}), this implies that $1 - X_j < \frac{1}{d_j}$ holds for every node $j$ in the graph. 
\end{remark} 

We now state the main result of this section. We show that degree-regular graphs achieve the highest $\frac{1}{n}\sum^n_{i=1} X_i$ over all graph topologies with the same average degree. 

\begin{theorem}\label{theorem:degreeheterogeneity}
Consider a {\it Total Effort} game on a graph under Assumptions~\ref{assumption:weightingfunction},~\ref{assumption:degreeheterogeneity} and~\ref{assumption:largeN}. Let $\frac{d_ic}{L} > w'(\mathbf{x}_{\min,w})$ for every player $i$, and let $d_{\avg} \triangleq \frac{1}{n} \sum^n_{i=1} d_i$ be the average of all extended neighborhood sizes. Then,
$$ \Phi(\mathbf{s}^*) \leq \frac{1}{n} \sum^n_{i=1} X_i \leq X_{\avg}, $$
where $\mathbf{s}^*$ is a PNE strategy profile and $X_{\avg} \in (\mathbf{x}_{\min,w},1]$ is such that $w'(X_{\avg}) = \frac{d_{\avg} b}{L}$.
\end{theorem}

\begin{proof}
Let the function $h: \mathbb{R}_+ \to [\mathbf{x}_{\min,w},\infty)$ be defined as the inverse of $w'(\cdot)$, i.e., $h(\frac{d_ic}{L}) \triangleq X_i$. From Assumptions~\ref{assumption:weightingfunction} and~\ref{assumption:degreeheterogeneity}, we know that $w'(x)$ is strictly increasing and is strictly convex for $x \in (\mathbf{x}_{\min,w},1]$. Therefore, the inverse function $h(\cdot)$ is strictly increasing and strictly concave in $d_i$ for $d_i \in \mathbb{R}$. From the strict concavity of $h(\cdot)$, we have $\frac{1}{n} \sum^n_{i=1} h\left(\frac{d_ic}{L}\right) \leq h\left(\frac{d_{\avg} c}{L}\right)$, which yields $\frac{1}{n} \sum^n_{i=1} X_i \leq X_{\avg}$. Equality holds when $d_i = d_{\avg}$ for every node $i$. 
\end{proof}

The above result states that graphs with identical node degrees have a larger worst case $\Phi(\mathbf{s}^*)$. Furthermore, as a graph becomes more dense, the bound on $\Phi(\mathbf{s}^*)$ grows with the average degree of the nodes. 
 
\subsection{Graphs with smallest average attack probability bound}

In this subsection, we answer the complementary question regarding graph topologies that have the smallest upper bound on $\Phi(\mathbf{s}^*)$.

In order to highlight the dependence of $X_i$ on $d_i$, we will use a slightly modified notation in the proof of the following result. In particular, we denote $X_i$ as $X_{d_i}$ for player $i$.

\begin{proposition}\label{proposition:star}
Consider {\it Total Effort} games with $n$ homogeneous players that satisfy Assumptions~\ref{assumption:weightingfunction},~\ref{assumption:degreeheterogeneity} and~\ref{assumption:largeN}. Let $\frac{dc}{L} > w'(\mathbf{x}_{\min,w})$ for $d = 2$. Then, among all connected graphs with $n$ nodes, the star graph achieves the smallest $\sum^n_{i=1} X_{d_i}$. 
\end{proposition}

\begin{proof} Recall from Remark~\ref{remark:increasingX} that $X_{d_i}$ is an increasing function of $d_i$. If the graph is not a tree, then we can remove a set of edges until the resulting subgraph is a tree. This reduces the neighborhood sizes and decreases $\sum^n_{i=1} X_{d_i}$.

It remains to show that among all trees, the star graph minimizes $\sum^n_{i=1} X_{d_i}$. Consider a tree that is not a star graph. Then, $\max^n_{i=1} d_i < n$. Consider a node $u$ with highest extended neighborhood size $d_u < n$. Since the graph is not a star, there must exist a leaf node $l$ which is connected to a node other than $u$. Let the neighbor of $l$ be denoted as $v$, with neighborhood size $d_v \leq d_u$. We argue that $\sum^n_{i=1} X_{d_i}$ decreases if we remove the edge between $l$ and $v$, and add an edge between $l$ and $u$. Under this operation, $X_{d_u}$ increases to $X_{d_u+1}$ for node $u$ and $X_{d_v}$ decreases to $X_{d_v-1}$ for node $v$. For all other nodes, $X_{d_i}$ remains unchanged. We compute the change in $\sum^n_{i=1} X_{d_i}$ as,
\begin{align}
& X_{d_u+1} - X_{d_u} + X_{d_v-1} - X_{d_v} \nonumber
\\ < & (X_{d_u} - X_{d_u-1}) - (X_{d_v} - X_{d_v-1}) \nonumber
\\ \leq & (X_{d_v} - X_{d_v-1}) - (X_{d_v} - X_{d_v-1}) = 0. 
\label{eq:firstorderdifferenceX}
\end{align}
The first inequality follows from Theorem~\ref{theorem:degreeheterogeneity} with two players having extended neighborhood sizes $d_u-1$ and $d_u+1$, i.e., $X_{d_u+1} + X_{d_u-1} < 2X_{d_u}$ for $d_u \geq 3$. The second inequality holds for $d_v \leq d_u$ following the strict concavity of $h(\cdot)$ (where $h(\frac{dc}{L}) = X_d$) as shown in Theorem~\ref{theorem:degreeheterogeneity}. Therefore, for any tree which is not a star graph, we can construct another tree which reduces $\sum^n_{i=1} X_{d_i}$. 
\end{proof}

The proof of Proposition~\ref{proposition:star} relies on the properties of the quantities $X_i$ which arise due to the nature of the behavioral weighting functions. In addition, the result that star graphs achieve the smallest security risk upper bound is an artifact of the {\it Total Effort} risk externalities, which assume that the attack on each node only depends on the average security investment in her neighborhood. In other words, the attacks are not targeted (in the sense of trying to disconnect the network). 

An interesting avenue for future work is to characterize network topologies that maximize more complex network value functions (such as those considered in~\cite{gueye2010design,schwartz2011network,cerdeiro2015contagion}) under behavioral probability weighting and targeted attacks. 

%% file: specificclasses.tex
In this section, we compare the effect of the intensity of probability weighting on $\Phi(\mathbf{s}^*)$ (the average equilibrium probability of successful attack) in degree-regular graphs. Analyzing degree-regular graphs not only isolates the effects of heterogeneity in degrees from the effects of the weighting function, but as we showed in the previous section, such graphs possess an {\it interior equilibrium} that upper bounds the $\Phi(\mathbf{s}^*)$ that arise in a broader class of graphs. 

In order to compare two different weighting functions, we use the parametric form~\eqref{eq:prelec} proposed by Prelec with parameter $\alpha \in (0,1]$. When $\alpha=1$, the weighting function is linear; as $\alpha$ decreases from $1$, the magnitudes of overweighting and underweighting increase. 

We consider two {\it Total Effort} games on a $(d-1)-$regular graph, i.e., the size of the extended neighborhood of the nodes is $d$. Let the cost parameters $c$ and $L$ be the same among the players across the two games. The first game has homogeneous players with weighting parameter $\alpha_1$, and the second game has homogeneous players with weighting parameter $\alpha_2$, with $\alpha_1 < \alpha_2 < \frac{dc}{L}$. In other words, the players in the first game have a more significant overweighting and underweighting of the true probabilities compared to the players in the second game. Let $X_1$ and $X_2$ be the solutions to the equations $w'_i(x) = \frac{dc}{L}$, $i \in \{1,2\}$, such that $X_i > \mathbf{x}_{\min,w}$. Here $w_i(x)$ is the Prelec weighting function \eqref{eq:prelec} with weighting parameter $\alpha_i$. From Proposition~\ref{proposition:degreeregular}, the attack probability at each node is equal to $X_i$ at the respective interior PNEs and $\Phi_i(\mathbf{s}^*) = X_i$. 

\begin{figure}[t]	
	\centerline{\includegraphics[width=6cm,height=4cm]{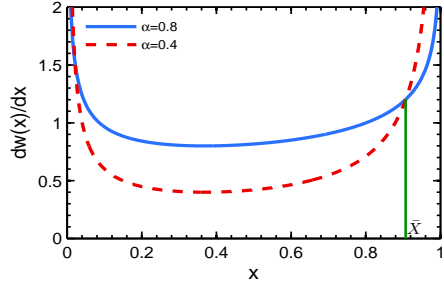}}
	\caption{$w'_i(x)$ for Prelec weighting functions with parameters $\alpha_1=0.4$ and $\alpha_2=0.8$. At $x = \bar{X}$, $w'_1(x) = w'_2(x)$.}
	\label{fig:twoalphadw}
\end{figure}

As we illustrate in Figure~\ref{fig:twoalphadw} for $\alpha_1 = 0.4$ and $\alpha_2 = 0.8$, $w'_1(x)$ is initially smaller than $w'_2(x)$ as $x$ starts to increase from $\frac{1}{e}$, until the quantity $x = \bar{X}$ (which depends on the values of $\alpha_1$ and $\alpha_2$) at which $w'_1(x) = w'_2(x)$. For $x > \bar{X}$, $w'_1(x) > w'_2(x)$. We state this formally in the following lemma. The proof is presented in Appendix~\ref{appendix:comparativestatics}. 

\begin{lemma}\label{lemma:twoalphafoc}
Consider two Prelec weighting functions $w_1(\cdot)$ and $w_2(\cdot)$ with parameters $\alpha_1$ and $\alpha_2$, respectively and let $\alpha_1 < \alpha_2$. Then there exists a unique $\bar{X} > \frac{1}{e}$ such that (1) $w'_1(\bar{X}) = w'_2(\bar{X})$, (2) for $x \in (\frac{1}{e},\bar{X})$, $w'_1(x) < w'_2(x)$, and (3) for $x \in (\bar{X},1)$, $w'_1(x) > w'_2(x)$.
\end{lemma}

Now we present the main result of this section, which follows from the above lemma. 

\begin{theorem}\label{theorem:compstat}
Let $X_i, i \in \{1,2\}$ be the true probability of successful attack at a node at the interior PNE of a ($d-1$)-regular graph where the nodes have a Prelec weighting function with parameter $\alpha_i, i \in \{1,2\}$, with $\alpha_1<\alpha_2$. Let $\bar{X}$ be the intersection point defined in Lemma~\ref{lemma:twoalphafoc} for $\alpha_1$ and $\alpha_2$. Then,
\begin{enumerate}
\item if $\frac{dc}{L} > w'(\bar{X})$, we have $\bar{X} < X_1 < X_2$,
\item otherwise, if $\alpha_2 < \frac{dc}{L} < w'(\bar{X})$, we have $\bar{X} > X_1 > X_2$, and
\item if $\frac{dc}{L} = w'(\bar{X})$, we have $\bar{X} = X_1 = X_2$.
\end{enumerate}
\end{theorem}

The above result shows that when $\frac{dc}{L}$ is such that the attack probability at a node at PNE is high enough (greater than $\bar{X}$), the players with substantial underweighting of probabilities (i.e., with smaller $\alpha$) view increased security investments to be highly beneficial in terms of reducing the perceived attack probabilities. As a result, the attack probability at a node at the PNE is smaller in the game with weighting parameter $\alpha_1$. 

On the other hand, when the attack probability is less than $\bar{X}$, the players with smaller $\alpha$ do not find the perceived reduction in attack probabilities to be sufficient to make a high investment. However players with weighting functions closer to linear observe greater perceived reduction in probability due to increased investment, and as a result, these players have smaller average attack probability at the interior equilibria. 

When we keep $\frac{c}{L}$ fixed and increase the neighborhood size $d$, then we eventually end up in the regime where the equilibrium attack probability is greater than $\bar{X}$. Thus, the expected fraction of nodes that are successfully attacked at equilibrium ($\Phi_i(\mathbf{s}^*)$) is smaller under behavioral probability weighting when the graph is sufficiently dense (i.e., has a large number of edges for a given number of nodes), and vice versa. 

%% file: weakestandbest.tex
The Nash equilibrium strategies in {\it Weakest Link} and {\it Best Shot} games have very special properties, as the security level at each neighborhood is determined by the investment of a single player (the one with the smallest and largest investment, respectively). Therefore, we first state the following results that characterize the security investment by an isolated player (i.e., $d=1$). The proofs of the following results are in Appendix~\ref{appendix:weakestandbest}. 

\begin{lemma}\label{lemma:opttransition}
Consider a weighting function that satisfies Assumption~\ref{assumption:weightingfunction}. Let $z > \mathbf{x}_{\min,w}$ be such that $w'(z) = \frac{w(z)}{z}$. Then i) $z$ exists and is unique, and ii) for $x > z$, $w'(x) > \frac{w(x)}{x}$. 
\end{lemma}

\begin{proposition}\label{proposition:singleplayer}\label{proposition:totaleffortsocialopt}
Let $z$ be as defined in Lemma~\ref{lemma:opttransition}. If $\frac{c}{L} < w'(z)$, the optimal investment of a single player is $s^* =1$. Otherwise, the optimal investment is $s^* = 1 - X$, where $w'(X) = \frac{c}{L}, X \in (\mathbf{x}_{\min,w},1]$.
\end{proposition}

We now analyze the PNE in {\it Weakest Link} games.

\begin{proposition}\label{proposition:weakestlink}
Consider a {\it Weakest Link} game on a connected graph with homogeneous players. Then at any PNE, all nodes make identical investments. If $\frac{c}{L} \geq w'(z)$, where $z$ is as defined in Lemma~\ref{lemma:opttransition}, then there is a continuum of pure Nash equilibria where the successful attack probabilities at the nodes are greater than or equal to $X$. When $w'(\mathbf{x}_{\min,w}) < \frac{c}{L} < w'(z)$, then there are additional equilibria (including the ones in the previous case) with attack probabilities close to $0$. 
\end{proposition}

\begin{proof}
Consider a node $v_1$ with security investment $s_1$, and let $\hat{s} < s_1$ be the smallest investment in her neighborhood. Then the attack probability on node $v_1$ is $f_i(s_i,\mathbf{s}_{\mathcal{N}(i)})  = 1- \hat{s} > 1 - s_1$. If node $v_1$ reduces her investment to $\hat{s}$, then her cost of investment decreases, while the attack probability remains unchanged. Therefore, at a PNE, the investment by any node must be equal to the minimum investment in her extended neighborhood. As a result, in connected graphs, all nodes must make identical security investments at a PNE. 

When $\frac{c}{L} \geq w'(z)$, Proposition~\ref{proposition:totaleffortsocialopt} states that a single node investing in isolation would prefer to invest $s^* = 1-X$. Now suppose all nodes have identical security investment $s \leq s^*$, i.e., the true attack probability at any node is $1-s \geq X$. Since for each player $w'(x) > \frac{c}{L}$ for $x > X$ (i.e., the marginal utility is positive), no player would unilaterally deviate to make a smaller investment. Therefore, any investment $s \leq 1-X$ by all the nodes would result in a PNE.

When $w'(\mathbf{x}_{\min,w}) < \frac{c}{L} < w'(z)$, the optimal investment of a single player is to invest $1$. Therefore, a strategy profile where each node invests $s = 1-\epsilon$ for sufficiently small $\epsilon > 0$ such that $\mathbb{E}u(s) > \mathbb{E}u(1-X)$, is a PNE, with attack probability $\epsilon$ at every node. For this strategy profile, all players have positive marginal utility, and prefer to invest $1-\epsilon$ over $1-X$ due to the continuity of the utility functions. These equilibria exist in addition to the set of equilibria with attack probabilities at least $X$. Note that any investment $s \in (V,X)$ by the nodes is not a PNE since $\frac{\partial \mathbb{E}u}{\partial s} < 0$ for $s \in (V,X)$. 
\end{proof}

Note that as long as the graph remains connected, its structure plays no role on the equilibrium investments and attack probabilities. The first part of the above result (i.e., identical investments by all players) holds for true expectation maximizers (players with $w(x)=x$) as well. The main differences with weighting functions are twofold. First, for large enough $\frac{c}{L}$, the only possible equilibrium with true expectation maximizers is when all players invest $0$ (Proposition~\ref{proposition:riskneutralresult} with $d=1$), while with probability weighting, there is a range of possible equilibrium investments, with resulting attack probabilities greater than $X$. Second, for $w'(\mathbf{x}_{\min,w}) < \frac{c}{L} < 1$, any investment by the players can give rise to a PNE for true expectation maximizers, while with probability weighting, there exist attack probabilities that can be supported at a PNE are either close to $0$ or at least $X$. Finally, when $\frac{c}{L} \leq w'(\mathbf{x}_{\min,w})$, $X$ is not defined, though the optimal investment by a single player is still $1$. In this case, any investment by the players can give rise to a PNE. 

We finally discuss the PNEs that arise in {\it Best Shot} games. 

\begin{proposition}\label{proposition:bestshotpne}
Consider a {\it Best Shot} network security game with homogeneous players. Then a strategy profile is a PNE if and only if there is a set of nodes who form a maximal independent set and invest according to Proposition~\ref{proposition:totaleffortsocialopt}, and all other nodes invest $0$. 
\end{proposition}

\begin{proof}
In a {\it Best Shot} game, the attack probability on a node is a function of the highest investment in her extended neighborhood. Suppose at a neighborhood there is a player making an investment in accordance with Proposition~\ref{proposition:totaleffortsocialopt}. Since this is the optimal investment a single node can make, her neighbors do not find investing more than this level (and thereby reducing their attack probabilities) profitable. Therefore, their optimal strategy is to invest $0$, as it eliminates their security investment cost $cs_i$, while the attack probability is unchanged. Therefore, any two nodes who make a nonzero investment must not be adjacent to each other. Furthermore, every node that is making a zero investment must have a neighbor who invests a nonzero amount. Therefore, the set of nodes making a nonzero investment must belong to a maximal independent set. The converse is also true; any maximal independent set with investments determined by Proposition~\ref{proposition:totaleffortsocialopt}, and all other nodes investing $0$ constitute a PNE.
\end{proof}

The independent set characterization holds for true expectation maximizers as well; nonlinear weighting functions change the level of investment. With true expectation maximizers, the nonzero investment level is at one of the boundary points, either $0$ or $1$ (Proposition~\ref{proposition:riskneutralresult} with $d=1$). With probability weighting, there are no equilibria that are entirely unprotected, and the nonzero equilibrium investment is at one of the interior solutions when $\frac{c}{L} > w'(z)$.

%% file: numerical.tex
We illustrate our theoretical characterizations of the impacts of network structure and probability weighting on the equilibrium investments in two numerical examples presented below. 

\begin{figure}[t] 
\centering
\begin{tikzpicture}[-,>=stealth',shorten >=1pt,auto,thick,main node/.style={circle,draw,minimum size = 0.6cm,inner sep=0pt}]
  
  \node[main node] (1) {$1$};
  \node[main node] (2) [below = 0.4cm of 1] {$2$};
  \node[main node] (3) [below = 0.4cm of 2] {$3$};
  \node[main node] (4) [right = 0.8cm of 1] {$4$};
  \node[main node] (5) [right = 1.5cm of 2] {$5$};
  \node[main node] (6) [right = 0.8cm of 3] {$6$};
  \node[main node] (7) [right = 1cm of 5] {$7$};
  \node[main node] (8) [right = 1cm of 7] {$8$};
  \node[main node] (9) [above = 0.4cm of 8] {$9$};
  \node[main node] (10) [below = 0.4cm of 8] {$10$};

  \path[every node/.style={font=\sffamily\small}]
    (1) edge (4)
    (2) edge (5)     
    (3) edge (6)
    (4) edge (5)
    (4) edge [bend right] (6)
    (5) edge (6)
    (4) edge (7)
    (5) edge (7)
    (6) edge (7)
    (7) edge (8)
    (8) edge (9)
    (8) edge (10);
\end{tikzpicture}
\caption{Graph topology analyzed in Example~\ref{ex:network}}
\label{fig:neighborhood}
\end{figure}
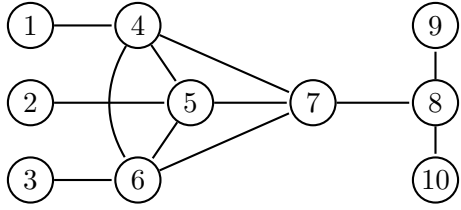

\begin{example}\label{ex:network}
Consider the graph shown in Figure~\ref{fig:neighborhood} with $n=10$ nodes. Each node is a decision maker with a Prelec weighting function with parameter $\alpha =  0.6$ and $\frac{c}{L}=0.45$. The extended neighborhood sizes range from $2$ to $5$. These parameters satisfy Assumptions~\ref{assumption:weightingfunction},~\ref{assumption:degreeheterogeneity} and~\ref{assumption:largeN}.

Under the Total Effort externalities, sequential best response dynamics converged to a PNE strategy profile in this example. In this PNE, the leaf nodes, $1,2,3,9,10$ have an investment $0.4095$, node $7$ invests $0.1442$ and nodes $4,5,6,8$ invest $0$. These investments satisfy \eqref{eq:neighborhoodaggregate}. Note that node $7$ has a larger neighborhood size compared to node $8$, yet it invests more than node $8$ at the PNE. This is in contrast with the equilibria that arise without probability weighting (Proposition~\ref{proposition:riskneutralresult}), where the equilibrium investments were either $0$ or $1$ and the investment of a node was at most that of a node with a smaller degree. Furthermore, the investments of the leaf nodes are larger than their neighbors, as shown in Proposition~\ref{proposition:extended_neighborhood}.

Under the Best Shot externality, a set of nodes which form a maximal independent set have a nonzero investment (Proposition~\ref{proposition:weakestlink}). In the graph shown in Figure~\ref{fig:neighborhood}, nodes $1,2,3,7,9,10$ form a maximal independent set. Another such set consists of nodes $4,2,3,8$. For the Prelec weighting function with $\alpha=0.6$, $w'(z)=0.8304$ for the value of $z$ defined in Lemma~\ref{lemma:opttransition}. Accordingly, if $\frac{c}{L} \leq 0.8304$, an equilibrium investment profile is when the nodes in a maximal independent set invest $1$, while other nodes invest $0$. On the other hand, if $\frac{c}{L} > 0.8304$, the equilibrium investment by the nodes is $1-X$, where $w'(X) = \frac{c}{L}, X \in (\frac{1}{e},1]$. 
\end{example}

In the above example, we discussed the effects of network structure on PNE investments in Total Effort and Best Shot games. The impact of the intensity of overweighting and underweighting of probabilities in degree-regular graphs is illustrated in the following example and corroborates our theoretical findings in Section~\ref{section:classesofgraphs}. 

\begin{example}\label{ex:weighting}
We consider two degree-regular graphs on $n=6$ nodes as shown in Figure~\ref{fig:weighting}. The nodes in the graph in Figure~\ref{fig:weighting3} (respectively, Figure~\ref{fig:weighting5}) have extended neighborhood sizes $d=3$ (respectively, $d=5$). We consider two types of players with Prelec weighting functions with parameters $\alpha_1 = 0.4$ and $\alpha_2 = 0.8$, respectively. The derivative of the weighting functions for such players was shown in Figure~\ref{fig:twoalphadw}. Let $\frac{c}{L}=0.3$ for all players.

\begin{figure}[t] 
\begin{subfigure}[b]{.49\linewidth}
\centering
\begin{tikzpicture}[-,>=stealth',shorten >=1pt,auto,thick,main node/.style={circle,draw,minimum size = 0.6cm,inner sep=0pt}]
  \node[main node] (1) {$1$};
  \node[main node] (2) [right = 0.5cm of 1] {$2$};
  \node[main node] (3) [below right = 0.5cm and 0.1cm of 2] {$3$};
  \node[main node] (4) [below left = 0.5cm and 0.1cm of 3] {$4$};
  \node[main node] (5) [left = 0.5cm of 4] {$5$};
  \node[main node] (6) [above left = 0.5cm and 0.1cm of 5] {$6$};

  \path[every node/.style={font=\sffamily\small}]
    (1) edge (2)
    (2) edge (3)     
    (3) edge (4)
    (4) edge (5)
    (5) edge (6)
    (6) edge (1);
    \end{tikzpicture}    
    \caption{$2$-regular graph on $6$ nodes}
    \label{fig:weighting3}
\end{subfigure}
\begin{subfigure}[b]{.49\linewidth} 
\centering 
\begin{tikzpicture}[-,>=stealth',shorten >=1pt,auto,thick,main node/.style={circle,draw,minimum size = 0.6cm,inner sep=0pt}]
  \node[main node] (1) {$1$};
  \node[main node] (2) [right = 0.5cm of 1] {$2$};
  \node[main node] (3) [below right = 0.5cm and 0.1cm of 2] {$3$};
  \node[main node] (4) [below left = 0.5cm and 0.1cm of 3] {$4$};
  \node[main node] (5) [left = 0.5cm of 4] {$5$};
  \node[main node] (6) [above left = 0.5cm and 0.1cm of 5] {$6$};

  \path[every node/.style={font=\sffamily\small}]
    (1) edge (2)
    (1) edge (3)
    (1) edge (5)
    (2) edge (3)
    (2) edge (4)
    (2) edge (6)     
    (3) edge (4)
    (4) edge (5)
    (4) edge (6)
    (5) edge (6)
    (5) edge (3)
    (6) edge (1);
    \end{tikzpicture}
    \caption{$4$-regular graph on $6$ nodes}
    \label{fig:weighting5}
\end{subfigure}    
\caption{Graph topologies analyzed in Example~\ref{ex:weighting}}
\label{fig:weighting}
\end{figure}
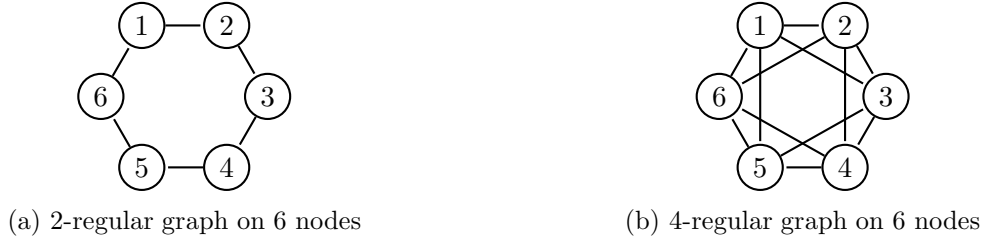

First we consider the $2$-regular graph shown in Figure~\ref{fig:weighting3}. When all the players have weighting functions with parameter $\alpha_1 = 0.4$, then the attack probability at every node at the interior equilibrium is equal to $0.8588$, while for players with parameter $\alpha_2 = 0.8$, the corresponding attack probability is $0.6912$. In this case, the equilibrium is more secure for the players whose weighting functions are closer to linear. In contrast, in the $4$-regular graph shown in Figure~\ref{fig:weighting5}, the attack probabilities at every node at the interior PNEs are equal to $0.9325$ and $0.9643$ for players with weighting function parameters $\alpha_1$ and $\alpha_2$, respectively.
\end{example}

As we noted in Theorem~\ref{theorem:compstat} and the succeeding discussion, the above example illustrates that the equilibrium is more secure for players with a substantial degree of overweighting and underweighting (captured by a smaller parameter value $\alpha$) when the graph is sufficiently dense.

%% file: appendix.tex
\appendix

\section{Proofs Pertaining to Equilibrium Characterization}
\label{appendix:existence}

\noindent {\bf Proof of Lemma~\ref{lemma:bestresponse}:} 

\begin{proof}
For ease of notation, we will drop the subscript $i$ since the proof holds for every player.

Let $\bar{s} \in [0, d-1]$ be the total security investment by the neighbors and $d$ be the size of the extended neighborhood of the player. Under Assumption~\ref{assumption:largeN}, the interval $\mathcal{X}(\bar{s})$ in~\eqref{eq:bestresponseinterval} falls into one of four different cases.

\textbf{Case 1:} $X < 1-\frac{1+\bar{s}}{d}$ (i.e., $\bar{s} < d(1-X)-1$)

In this case, the interval $\mathcal{X}(\bar{s})$ lies to the right of $X$. Therefore, at any attack probability $x \in \mathcal{X}(\bar{s})$, $\frac{L}{d}w'(x) > c$ (from Figure~\ref{fig:interiorfoc}). Thus, $\frac{\partial \mathbb{E}u}{\partial s}\rvert_{s = x} > 0$ and consequently, $b(\bar{s}) = 1$. 

\textbf{Case 2:} $1-\frac{1+\bar{s}}{d} \leq X \leq 1-\frac{\bar{s}}{d}$

In this case, $X \in \mathcal{X}(\bar{s})$, and therefore, the player has a feasible investment strategy $s^* = d(1-X) - \bar{s}$ at which the first order condition is satisfied with equality. For any investment $y < s^*$, we have the resulting attack probability $1-\frac{y + \bar{s}}{d} > X$ and $\frac{\partial \mathbb{E}u}{\partial s} = \frac{L}{d}w'(X)-c > 0$. As a result, no value of $y < s^*$ would satisfy the first order necessary condition of optimality. On the other hand, for any $y > s^*$, we have $1-\frac{y + \bar{s}}{d} < X$. However, from our assumption that $X > \frac{1}{d} + V$, we would have $1-\frac{y + \bar{s}}{d} > V$. Therefore, $\frac{\partial \mathbb{E}u}{\partial s} < 0$ for any $y > s^*$. As a result, $x = s^*$ is the only candidate for optimal investment, and it also satisfies the second order sufficient condition since $w''(X) > 0$ as $X > \mathbf{x}_{\min,w}$. Since the optimal solution $s^*$ must have the property that $1-\frac{s^*+\bar{s}}{d} = X$, it is continuous and linearly decreasing in $\bar{s}$, with boundary values at $1$ and $0$ for $\bar{s} = d(1-X) - 1$ and $\bar{s} = d(1-X)$, respectively. 

In the remaining two cases, we have $\bar{s} > d(1-X)$.

\textbf{Case 3:} $V < 1-\frac{1+\bar{s}}{d}$ and $X > 1-\frac{\bar{s}}{d}$

In this case, the interval $\mathcal{X}(\bar{s})$ lies in the region between $V$ and $X$. Therefore, for any true attack probability $x \in \mathcal{X}(\bar{s})$, $\frac{\partial \mathbb{E}u}{\partial s} < 0$. As a result, $b(\bar{s}) = 0$.

\textbf{Case 4:} $1-\frac{1+\bar{s}}{d} \leq V \leq 1-\frac{\bar{s}}{d}$

In this case, there are three candidate solutions for utility maximization, $s = 1$, $s = d(1- V)-\bar{s}$ and $s = 0$. We have $1-\frac{s+\bar{s}}{d}$ as the true attack probability resulting from the strategies of the players. 

First we show that the player would always prefer to invest $1$ over investing $s^* = d(1- V)-\bar{s}$. We denote the resulting attack probability while investing $1$ as $Y_1 = 1-\frac{1+ \bar{s}}{d}$. Note that $V - Y_1 = \frac{1}{d}(1-s^*)$. Now using~\eqref{eq:weightedutility} we compute
\begin{align*}
& \mathbb{E}u(1,\mathbf{s}_{\mathcal{N}}) - \mathbb{E}u(s^*,\mathbf{s}_{\mathcal{N}}) 
\\ = & L(w(V) - w(Y_1)) - c(1-s^*)
\\ = & L(w(V) - w(Y_1)) - cd(V-Y_1)
\\ = & L(V-Y_1) \left[ \frac{w(V) - w(Y_1)}{(V-Y_1)} - \frac{dc}{L}\right]
\\ > & L(V-Y_1) \left[w'(V) - \frac{dc}{L}\right] = 0,
\end{align*}
where the inequality is due to the fact that $w(x)$ is strictly concave for $x \in [0,\mathbf{x}_{\min,w})$, and $Y_1 < V < \mathbf{x}_{\min,w}$. Therefore, the player will always prefer the boundary solution $s=1$ to the potential interior solution $s^*$.

This leads to the possibility that the best response might have a discontinuous jump from $0$ to $1$ at some value of $\bar{s}$ in this region. However, we show that under the second and third conditions of Assumption~\ref{assumption:largeN}, the player would always prefer to invest $0$ over investing $1$. We compute
\begin{align*}
& \mathbb{E}u(1,\mathbf{s}_{\mathcal{N}}) - \mathbb{E}u(0,\mathbf{s}_{\mathcal{N}}) 
\\ = & L\left[w\left(1-\frac{\bar{s}}{d}\right) - w\left(1-\frac{1+\bar{s}}{d}\right)\right] - c
\\ = & L\left[w\left(\lambda+\frac{1}{d}\right) - w(\lambda)\right] - c \leq L\left[w\left(\frac{1}{d}\right) - w(0)\right] - c,
\end{align*}
where $\lambda = 1-\frac{1+\bar{s}}{d}$. The last inequality follows because the function $l_1(\lambda) \triangleq w(\lambda+\frac{1}{d}) - w(\lambda)$ is a strictly decreasing function of $\lambda$ for $\lambda \in [0,V]$. Indeed, $l'_1(\lambda) = w'(\lambda+\frac{1}{d}) - w'(\lambda) < 0$, as $w'(\lambda) \geq w'(V) = \frac{dc}{L}$, and $w'(\lambda + \frac{1}{d}) < \frac{dc}{L}$, as $V \leq \lambda + \frac{1}{d} < X$. Therefore, if $w(\frac{1}{d}) < \frac{c}{L}$, the player would always prefer to invest $0$ over investing $1$, regardless of the value of $\bar{s}$ (including when $\bar{s} = d-1$). 
\end{proof}

\noindent {\bf Proof of Proposition~\ref{proposition:totaleffortNEheterounique}:} 

\begin{proof}
Without loss of generality, let players be ordered such that $X_1 \leq X_2 \leq X_3 \leq \ldots \leq X_n$, where $X_i$ is the largest solution to $w'_i(x) = \frac{nc_i}{L_i}$. When there is no solution to $w'_i(x) = \frac{nc_i}{L_i}$, i.e., $\min_{x \in [0,1]} w'(x) > \frac{nc_i}{L_i}$ for a player, then we define $X_i = 0$ for that player. When $X_n = 0$, then all players investing $1$ is the only PNE.

The true attack probability at the PNE must be at least $0$. When $X_1 > 0$, no objective attack probability $0 \leq X < X_1$ would be a PNE, since there would always exist a player with positive investment who would prefer to invest $0$. 

Now suppose there are two PNEs with different probabilities of successful attack $X^*$ and $Y^*$, with $X^*<Y^*$. Consider the strategy profile with the smaller attack probability $X^*$. Note that we have ruled out the possibility of $X^* < X_1$ above. There are two exhaustive cases:  either $X_l < X^* < X_{l+1}$ for some player $l$, or $X_l = X^*$ for some player $l$. 

Let $X_l < X^* < X_{l+1}$ for some player $l$. By the definition of the quantities $X_i$, we have $w'_i(X^*) < \frac{nc_i}{L_i}$ for $i \in \{l+1,\ldots,n\}$, and therefore, $s^*_i = 0$ for $i \in \{l+1,\ldots,n\}$. Similarly, $w'_i(X^*) > \frac{nc_i}{L_i}$ and $s^*_i = 1$ for $i \in \{1,\ldots,l\}$. In this case, $X^* = 1 - \frac{l}{n}$. Now at the second PNE with true attack probability $Y^* > X^*$, the players in $\{1,\ldots,l\}$ would continue to invest $1$, with the possibility of more players investing nonzero amounts if $Y^* \geq X_{l+1}$. But then the true attack probability would decrease from $X^*$, contradicting the assumption that $Y^* > X^*$. The proof of the case where $X^* = X_l$ for some player $l$ follows identical arguments.
\end{proof}
 
\section{Linear Complementarity Problem Formulation}
\label{appendix:lcp}

We can exploit the structure of the best response as presented in \eqref{eq:neighborhoodaggregate} to represent the Nash equilibrium strategy profile as a solution to a Linear Complementarity Problem (LCP).\footnote{Formally, given a vector $q \in \mathbb{R}^n$, and a matrix $M \in \mathbb{R}^{n \times n}$, the $LCP(q,M)$ is the problem of finding a solution vector $z \in \mathbb{R}^n$ such that i) $z \geq 0$, ii) $q+Mz \geq 0$, and iii) $z^\mathsf{T}(q+Mz) = 0$.}  

\begin{proposition}
Consider a total effort game on a graph with $n$ vertices with adjacency matrix $A$, and let $\frac{d_ic_i}{L_i} > w'_i(\mathbf{x}_{\min,w_i})$ for every player $i$. Then, under Assumptions~\ref{assumption:weightingfunction} and \ref{assumption:largeN}, a strategy profile is a Nash equilibrium of this game if and only if it is a solution of the $LCP(\mathbf{q},\mathbf{M})$, where $\mathbf{q} \in \mathbb{R}^{2n}$ and $\mathbf{M} \in \mathbb{R}^{2n \times 2n}$ are given by
\begin{align*}
\mathbf{q} =  \begin{bmatrix} -\mathbf{d}(1-\mathbf{X}_2) \\  \mathbf{1}_{n \times 1} \end{bmatrix}   \text{and }  \mathbf{M} = \begin{bmatrix} A+I_n \quad I_n \\  -I_n \quad \quad \mathbf{0}_{n \times n} \end{bmatrix},
\end{align*}
where the $j$th entry of $\mathbf{q}$ is $d_j(1-X^j_2)$ for $j \in \{1,2,\ldots,n\}$. 
\label{proposition:LCPdefinition}
\end{proposition}

\begin{proof}
The solution to the $LCP(\mathbf{q},\mathbf{M})$ is a vector $\mathbf{z} \in \mathbb{R}^{2n}$ which we can write as $\mathbf{z} = \begin{bmatrix} \mathbf{s} \\  \mathbf{\mu} \end{bmatrix}$, where $\mathbf{s}$ is the vector of investments by the players (strategy profile) and $\mathbf{\mu}$ is a set of variables which assume nonzero values when a corresponding investment $s_i = 1$. 

From the definition of a LCP, the solution $\mathbf{z} \geq 0$, i.e., $\mathbf{s} \geq 0$ and $\mathbf{\mu} \geq 0$. Secondly, from $\mathbf{q} + \mathbf{M}\mathbf{z} \geq 0$, for each player $i$, we have
\begin{align*}
& s_i + \sum_{j \in \mathcal{N}(i)} s_j + \mu_i \geq d_i (1-X^i_2), \text{\quad and }
\\ & - s_i + 1 \geq 0.
\end{align*}
The second inequality ensures that $s_i \leq 1$. Therefore, the investment by any player at any solution of the LCP is feasible, i.e., $s_ i \in [0,1]$. 

Finally, $\mathbf{z}^\mathsf{T}(\mathbf{q}+\mathbf{M}\mathbf{z}) = 0$ gives us
\begin{align*}
& s_i (s_i + \sum_{j \in \mathcal{N}(i)} s_j + \mu_i - d_i (1-X^i_2)) = 0, \text{\quad and }
\\ & \mu_i(- s_i + 1) = 0,
\end{align*}
for every node $i$. As a result, when $s_i \in (0,1)$, we have $\mu_i = 0$, and therefore, the investments satisfy the second part of equation~\eqref{eq:neighborhoodaggregate}. When $s_i = 0$, we also have $\mu_i = 0$, and the resulting investments satisfy the third part of equation~\eqref{eq:neighborhoodaggregate}. Finally, when $s_i = 1$, we have $\mu_i > 0$ such that the first part of equation~\eqref{eq:neighborhoodaggregate} holds. This concludes the proof. 
\end{proof}

A comprehensive discussion on LCPs and different solutions algorithms can be found in~\cite{cottle1992linear}. The structure of the LCP often determines the performance of different algorithms. We show that the LCP defined in Proposition~\ref{proposition:LCPdefinition} satisfies certain properties, as proven in Proposition~\ref{proposition:lemke}, that guarantee the convergence of the Lemke's pivotal method to converge to a solution (i.e., a PNE strategy profile), if the problem is non degenerate. The convergence result is due to~\cite{cottle1992linear}.  

\begin{proposition}
For the $LCP(\mathbf{q},\mathbf{M})$ defined in Proposition~\ref{proposition:LCPdefinition},
\begin{enumerate}
\item the matrix $\mathbf{M}$ is copositive and 
\item $\mathbf{q} \in SOL(\mathbf{0},\mathbf{M})^*$.
\end{enumerate}
\label{proposition:lemke}
\end{proposition}

\begin{proof}
For the proof of the first statement, consider any vector $\mathbf{x}^\mathsf{T} = [\mathbf{x_1} \quad \mathbf{x_2}]^\mathsf{T}$ where $\mathbf{x_i} \geq \mathbf{0}$, for $i = 1,2$. Then we have,
\begin{align*}
\mathbf{x}^\mathsf{T} \mathbf{M} \mathbf{x} & = [\mathbf{x_1} \quad \mathbf{x_2}]^\mathsf{T} \begin{bmatrix} A\mathbf{x_1} + \mathbf{x_1}  + \mathbf{x_2} \\  - \mathbf{x_1} \end{bmatrix}
\\ & = \mathbf{x}^\mathsf{T}_1 A \mathbf{x_1} + \mathbf{x}^\mathsf{T}_1\mathbf{x_1} \geq 0.
\end{align*}

For the second part, consider any solution $\mathbf{y} = \begin{bmatrix} \mathbf{y}_1 \\ \mathbf{y}_2 \end{bmatrix} $ of the $LCP(\mathbf{0},\mathbf{M})$. Then, we must have $\mathbf{y} \geq \mathbf{0}$, and $\mathbf{M}\mathbf{y} = \begin{bmatrix} (A+I_n)\mathbf{y}_1 + \mathbf{y}_2 \\ -\mathbf{y}_1 \end{bmatrix} \geq \mathbf{0}$. As a result, we must have $\mathbf{y}_1 = \mathbf{0}$.

For any vector $\mathbf{y}^\mathsf{T} = \begin{bmatrix} \mathbf{y}^\mathsf{T}_1 \quad \mathbf{y}^\mathsf{T}_2 \end{bmatrix}$, with $\mathbf{y}_1 = \mathbf{0}$ and $\mathbf{y}_2 \geq \mathbf{0}$, we have $\mathbf{q}^\mathsf{T} \mathbf{y} = \mathbf{1}_{n \times 1}^\mathsf{T} \mathbf{y} \geq 0$. Thus, $\mathbf{q} \in SOL(\mathbf{0},\mathbf{M})^*$.
\end{proof}

\section{Proofs Pertaining to the Effects of Network Structure}
\label{section:appendixDegreeHeterogeneity}

\noindent {\bf Proof of Lemma~\ref{lemma:doneminusx}:} 

\begin{proof}
We denote the inverse of $w'(\cdot)$ as the function $h(\cdot)$, i.e., $h: \mathbb{R}_+ \to [\mathbf{x}_{\min,w},\infty)$ such that $h(\frac{dc}{L}) = X$. Now, consider the function
\begin{align*}
\phi(d) & \triangleq d(1-X) = d\left(1-h\left(\frac{dc}{L}\right)\right)
\\ \implies \phi'(d) &= 1 - h\left(\frac{dc}{L}\right) - d\left[h\left(\frac{dc}{L}\right)\right]^{'}
\\ &= 1-h\left(\frac{dc}{L}\right) -\frac{dc}{L} \frac{1}{w''(h\left(\frac{dc}{L}\right))}
\\ & = 1 - X - \frac{dc}{L} \frac{1}{w''(X)}.
\end{align*} 
The second inequality follows from differentiating the inverse of the $w'$ function. Now for $x \in (\mathbf{x}_{\min,w},1)$, the weighting function $w$ satisfies
\begin{align*}
& \frac{w''(x)}{w'(x)} < \frac{1}{1-x} 
\\ \implies & (1-X) < \frac{w'(X)}{w''(X)} = \frac{dc}{L} \frac{1}{w''(X)} \implies \phi'(d) < 0. 
\end{align*}
Since $\phi(d)$ is strictly decreasing in $d$, we have $d(1-X)$ strictly decreasing when $d$ is interpreted as the size of the extended neighborhood of a node.
\end{proof}

\noindent {\bf Proof of Proposition~\ref{proposition:upperbound}:} 

\begin{proof}
We first show that when $d_i(1-X_i) < 1$, the attack probability at node $i$ is at most $X_i$ at a PNE. Recall from Lemma~\ref{lemma:bestresponse} that the best response of player $i$ is $b_i(\bar{s}_{-i}) = 1$ when the aggregate investment by her neighbors $\bar{s}_{-i} \leq d_i(1-X_i) - 1$. When $d_i(1-X_i) < 1$, the above condition is not satisfied since $\bar{s}_{-i} \geq 0$. As a result, the investment of player $i$ at any PNE lies in $[0,1)$. When $\bar{s}_{-i} \geq d_i(1-X_i)$, the investment by node $i$ is $0$, and the resulting attack probability is at most $X_i$. Otherwise, the sum of the investments by node $i$ and her neighbors satisfy the first order condition \eqref{eq:first_order} with equality, in which case the resulting attack probability is exactly $X_i$. 

The upper bound on average attack probability follows by averaging $X_i$'s over all nodes. At an interior PNE, each player experiences an attack probability exactly equal to $X_i$ and therefore the bound holds with equality. 
\end{proof}

\section{Proofs Pertaining to the Impact of Weighting Function}
\label{appendix:comparativestatics}

We first state and prove the following lemma whose result will be useful in the proof of Lemma~\ref{lemma:twoalphafoc}. 

\begin{lemma}\label{lemma:gmonotonicity}
The function 
$$
g(x) \triangleq (-\ln(x))^{\alpha_2 - \alpha_1} \exp((-\ln(x))^{\alpha_1} - (-\ln(x))^{\alpha_2}),
$$ 
with $\alpha_1 < \alpha_2 < 1$ is strictly decreasing for $x \in [\frac{1}{e},1]$.
\end{lemma}

\begin{proof}
We compute
\begin{align*}
g'(x) & = \exp((-\ln(x))^{\alpha_1} - (-\ln(x))^{\alpha_2}) \times \\ 
& \quad \left[ (\alpha_1(-\ln(x))^{\alpha_1-1} - \alpha_2(-\ln(x))^{\alpha_2-1}) \frac{-1}{x} \times \right. \\ & \qquad \left. (-\ln(x))^{\alpha_2 - \alpha_1} - (\alpha_2 - \alpha_1) (-\ln(x))^{\alpha_2 - \alpha_1-1} \frac{1}{x} \right]
\\ & = \frac{1}{x} \exp((-\ln(x))^{\alpha_1} - (-\ln(x))^{\alpha_2}) (-\ln(x))^{\alpha_2 - \alpha_1-1} \\ 
& \quad \times \left[-(\alpha_1(-\ln(x))^{\alpha_1} + \alpha_1 + \alpha_2(-\ln(x))^{\alpha_2} - \alpha_2\right].
\end{align*}
When $x > \frac{1}{e}$, $(-\ln(x)) < 1$ and thus $1 > (-\ln(x))^{\alpha_1} > (-\ln(x))^{\alpha_2}$. As a result, $1 - (-\ln(x))^{\alpha_1} < 1 - (-\ln(x))^{\alpha_2}$. Since $\alpha_1 < \alpha_2$, this implies $\alpha_1(-\ln(x))^{\alpha_1} - \alpha_1 > \alpha_2(-\ln(x))^{\alpha_2} - \alpha_2$. Therefore, $g'(x) < 0$ for $x \in [\frac{1}{e},1]$.
 \end{proof}

\noindent {\bf Proof of Lemma~\ref{lemma:twoalphafoc}:}

\begin{proof}
The first derivative of the Prelec weighting function is given by $w'(x) = w(x) \frac{\alpha}{x} (-\ln(x))^{\alpha-1}$.  Therefore, if at a given $x$, $w'_1(x) = w'_2(x)$,  we have
\begin{align*}
& w_1(x) \alpha_1 (-\ln(x))^{\alpha_1} = w_2(x) \alpha_2 (-\ln(x))^{\alpha_2}
\\ \implies & \frac{\alpha_1}{\alpha_2} = \frac{w_2(x) (-\ln(x))^{\alpha_2}}{w_1(x) (-\ln(x))^{\alpha_1}}
\\ \implies & \frac{\alpha_1}{\alpha_2} = (-\ln(x))^{\alpha_2 - \alpha_1} \exp((-\ln(x))^{\alpha_1} - (-\ln(x))^{\alpha_2})  
\\ \implies & \frac{\alpha_1}{\alpha_2} = g(x).
\end{align*}
From the definition of $g(x)$, $g(\frac{1}{e}) = 1 > \frac{\alpha_1}{\alpha_2}$. Furthermore, as $x \to 1$, $g'(x) \to -\infty$. As a result, $g(x)$ becomes smaller than $\frac{\alpha_1}{\alpha_2}$ for some $x < 1$. Thus there exists $\bar{X}$ at which $w'_1(x) = w'_2(x)$. The uniqueness of $\bar{X}$ follows from the strict monotonicity of $g(x)$ as proved in Lemma~\ref{lemma:gmonotonicity}. 

In order to prove the second and third parts of the lemma, it suffices to show that $w_1''(\bar{X}) > w_2''(\bar{X})$. Therefore, we compute
\begin{align*}
w''(x) = \frac{w'(x)}{-x \ln(x)} [1+\ln(x) + \alpha ((-\ln(x))^\alpha - 1)].
\end{align*}
From the previous discussion in Lemma~\ref{lemma:gmonotonicity}, $\alpha_1(-\ln(x))^{\alpha_1} - \alpha_1 > \alpha_2(-\ln(x))^{\alpha_2} - \alpha_2$, and $w''(x) > 0$ for $x > \frac{1}{e}$. Therefore at $\bar{X}$, $w_1''(\bar{X}) > w_2''(\bar{X})$. This concludes the proof.
 \end{proof}

\section{Proofs Pertaining to Weakest Link and Best Shot Games}
\label{appendix:weakestandbest}

\noindent {\bf Proof of Lemma~\ref{lemma:opttransition}:} 

\begin{proof}
Consider the function $l_2(x) \triangleq w'(x) - \frac{w(x)}{x}$ for $x \in [\mathbf{x}_{\min,w},1)$. At $x = \mathbf{x}_{\min,w}$, $l_2(x) = - \frac{w(\mathbf{x}_{\min,w})}{\mathbf{x}_{\min,w}} < 0$. Furthermore, $l_2(1-\epsilon) \to \infty$ as $\epsilon \to 0$. Therefore, $l_2(x)$ must have a root for $x \in [\mathbf{x}_{\min,w},1)$.

Suppose there exist $z_1,z_2$ with $z_1 < z_2$ such that $w'(z_i) = \frac{w(z_i)}{z_i}, i \in \{1,2\}$. Since $w(x)$ is strictly convex for $x \in (\mathbf{x}_{\min,w},1]$, we have
\begin{align*}
w(z_1) &> w(z_2) + w'(z_2) (z_1-z_2) = z_1 w'(z_2)
\\ \implies \frac{w(z_1)}{z_1} &> w'(z_2) > w'(z_1),
\end{align*}
which is a contradiction. Thus, $z$ is unique.

For the second part, suppose that there exists an $x > z$ at which $w'(x) \leq \frac{w(x)}{x}$. Again from the convexity of $w$, we have
\begin{align*}
w(z) &> w(x) + w'(x) (z-x) \geq zw'(x)
\\ \implies \frac{w(z)}{z} &> w'(x) > w'(z),
\end{align*}
which contradicts the definition of $z$. 
\end{proof}

\noindent {\bf Proof of Proposition~\ref{proposition:singleplayer}:} 

\begin{proof}
When there is a single player investing in isolation, we have three candidate solutions for utility maximization, $s_1 = 1-V$, $s_2 =1-X$ or $s_3 = 1$. Here $X$ and $V$ are solutions to $w'(x) = \frac{c}{L}$, as defined in Section~\ref{section:nashequilibriumexistence}. Note that since $\frac{c}{L}$ is finite, we have $\frac{\partial \mathbb{E}u}{\partial s}>0$ at $s=0$, so investing $0$ in security is not a utility maximizer. 

From the analysis in Case 4 of Lemma~\ref{lemma:bestresponse} with $\bar{s}=0$ and $d=1$, we have $\mathbb{E}u(1) > \mathbb{E}u(1-V)$. Therefore, between the potential interior solution $s_1 = 1-V$ that satisfies the first order condition, and the boundary solution $s_3 = 1$, the player always prefers the boundary solution. 

Now, to compare the utilities at the solutions $s_2$ and $1$, we compute  
\begin{align*}
\mathbb{E}u(1) - \mathbb{E}u(s_2) &= Lw(1-s_2) - c(1-s_2)
\\ & =  L(1-s_2) \left[ \frac{w(1-s_2)}{1-s_2} - w'(1-s_2)\right],
\end{align*}
where $\frac{c}{L} = w'(1-s_2)$. Thus, from Lemma~\ref{lemma:opttransition}, when $w'(1-s_2) = \frac{c}{L} > w'(z)$, then, $1-s_2 > z$ and the player prefers to invest $s_2$. Otherwise, the optimal investment is $1$.
\end{proof} 

%% file: root_old.bbl
\begin{thebibliography}{37}
\providecommand{\natexlab}[1]{#1}
\providecommand{\url}[1]{\texttt{#1}}
\expandafter\ifx\csname urlstyle\endcsname\relax
  \providecommand{\doi}[1]{doi: #1}\else
  \providecommand{\doi}{doi: \begingroup \urlstyle{rm}\Url}\fi

\bibitem[Amin et~al.(2013)Amin, Schwartz, and Sastry]{amin2013security}
Saurabh Amin, Galina~A Schwartz, and S~Shankar Sastry.
\newblock Security of interdependent and identical networked control systems.
\newblock \emph{Automatica}, 49\penalty0 (1):\penalty0 186--192, 2013.

\bibitem[Booij et~al.(2010)Booij, Van~Praag, and Van
  De~Kuilen]{booij2010parametric}
Adam~S Booij, Bernard~MS Van~Praag, and Gijs Van De~Kuilen.
\newblock A parametric analysis of prospect theory's functionals for the
  general population.
\newblock \emph{Theory and Decision}, 68\penalty0 (1-2):\penalty0 115--148,
  2010.

\bibitem[Bramoull{\'e} et~al.(2014)Bramoull{\'e}, Kranton, and
  D'amours]{bramoulle2014strategic}
Yann Bramoull{\'e}, Rachel Kranton, and Martin D'amours.
\newblock Strategic interaction and networks.
\newblock \emph{The American Economic Review}, 104\penalty0 (3):\penalty0
  898--930, 2014.

\bibitem[Camerer et~al.(2011)Camerer, Loewenstein, and
  Rabin]{camerer2011advances}
Colin~F Camerer, George Loewenstein, and Matthew Rabin.
\newblock \emph{Advances in behavioral economics}.
\newblock Princeton University Press, 2011.

\bibitem[Cerdeiro et~al.(2015)Cerdeiro, Dziubinski, and
  Goyal]{cerdeiro2015contagion}
Diego Cerdeiro, Marcin Dziubinski, and Sanjeev Goyal.
\newblock Contagion risk and network design.
\newblock \emph{Working paper}, 2015.

\bibitem[Christin(2011)]{christin2011network}
Nicolas Christin.
\newblock Network security games: combining game theory, behavioral economics,
  and network measurements.
\newblock In \emph{Decision and Game Theory for Security}, pages 4--6.
  Springer, 2011.

\bibitem[Christin et~al.(2011)Christin, Egelman, Vidas, and
  Grossklags]{christin2012s}
Nicolas Christin, Serge Egelman, Timothy Vidas, and Jens Grossklags.
\newblock It’s all about the {B}enjamins: An empirical study on incentivizing
  users to ignore security advice.
\newblock In \emph{Financial Cryptography and Data Security}, pages 16--30.
  Springer, 2011.

\bibitem[Cottle et~al.(1992)Cottle, Pang, and Stone]{cottle1992linear}
Richard~W Cottle, Jong-Shi Pang, and Richard~E Stone.
\newblock \emph{The linear complementarity problem}, volume~60.
\newblock SIAM, 1992.

\bibitem[Galeotti et~al.(2010)Galeotti, Goyal, Jackson, Vega-Redondo, and
  Yariv]{galeotti2010network}
Andrea Galeotti, Sanjeev Goyal, Matthew~O Jackson, Fernando Vega-Redondo, and
  Leeat Yariv.
\newblock Network games.
\newblock \emph{The Review of Economic Studies}, 77\penalty0 (1):\penalty0
  218--244, 2010.

\bibitem[Garg and Camp(2013)]{garg2013heuristics}
Vaibhav Garg and Joseph Camp.
\newblock Heuristics and biases: Implications for security design.
\newblock \emph{IEEE Technology and Society Magazine}, 32\penalty0
  (1):\penalty0 73--79, 2013.

\bibitem[Gharesifard et~al.(2016)Gharesifard, Touri, Ba{\c{s}}ar, and
  Shamma]{gharesifard2015convergence}
Bahman Gharesifard, Behrouz Touri, Tamer Ba{\c{s}}ar, and Jeff Shamma.
\newblock On the convergence of piecewise linear strategic interaction dynamics
  on networks.
\newblock \emph{IEEE Transactions on Automatic Control}, 61\penalty0
  (6):\penalty0 1682--1687, 2016.

\bibitem[Gonzalez and Wu(1999)]{gonzalez1999shape}
Richard Gonzalez and George Wu.
\newblock On the shape of the probability weighting function.
\newblock \emph{Cognitive Psychology}, 38\penalty0 (1):\penalty0 129--166,
  1999.

\bibitem[Grossklags and Johnson(2009)]{grossklags2009uncertainty}
Jens Grossklags and Benjamin Johnson.
\newblock Uncertainty in the weakest-link security game.
\newblock In \emph{Game Theory for Networks}, pages 673--682, 2009.

\bibitem[Grossklags et~al.(2008)Grossklags, Christin, and
  Chuang]{grossklags2008security}
Jens Grossklags, Nicolas Christin, and John Chuang.
\newblock Security and insurance management in networks with heterogeneous
  agents.
\newblock In \emph{9th ACM Conference on Electronic Commerce}, pages 160--169,
  2008.

\bibitem[Gueye et~al.(2010)Gueye, Walrand, and Anantharam]{gueye2010design}
Assane Gueye, Jean~C Walrand, and Venkat Anantharam.
\newblock Design of network topology in an adversarial environment.
\newblock In \emph{Decision and Game Theory for Security}, pages 1--20.
  Springer, 2010.

\bibitem[Hirshleifer(1983)]{hirshleifer1983weakest}
Jack Hirshleifer.
\newblock From weakest-link to best-shot: The voluntary provision of public
  goods.
\newblock \emph{Public Choice}, 41\penalty0 (3):\penalty0 371--386, 1983.

\bibitem[Hota and Sundaram(2015)]{hota2015interdependent}
Ashish~R Hota and Shreyas Sundaram.
\newblock Interdependent security games under behavioral probability weighting.
\newblock In \emph{Decision and Game Theory for Security}, pages 150--169.
  Springer, 2015.

\bibitem[Hota et~al.(2016)Hota, Garg, and Sundaram]{hota2014fragility}
Ashish~R Hota, Siddharth Garg, and Shreyas Sundaram.
\newblock Fragility of the commons under prospect-theoretic risk attitudes.
\newblock \emph{Games and Economic Behavior}, 98:\penalty0 135--164, 2016.

\bibitem[Jiang et~al.(2011)Jiang, Anantharam, and Walrand]{jiang2011bad}
Libin Jiang, Venkat Anantharam, and Jean Walrand.
\newblock How bad are selfish investments in network security?
\newblock \emph{IEEE/ACM Transactions on Networking}, 19\penalty0 (2):\penalty0
  549--560, 2011.

\bibitem[Johnson et~al.(2010)Johnson, Grossklags, Christin, and
  Chuang]{johnson2010are}
Benjamin Johnson, Jens Grossklags, Nicolas Christin, and John Chuang.
\newblock Are security experts useful? {B}ayesian {N}ash equilibria for network
  security games with limited information.
\newblock In \emph{ESORICS}, pages 588--606. Springer, 2010.

\bibitem[Kunreuther and Heal(2003)]{kunreuther2003interdependent}
Howard Kunreuther and Geoffrey Heal.
\newblock Interdependent security.
\newblock \emph{Journal of Risk and Uncertainty}, 26\penalty0 (2-3):\penalty0
  231--249, 2003.

\bibitem[Laszka et~al.(2014)Laszka, Felegyhazi, and Buttyan]{laszka2014survey}
Aron Laszka, Mark Felegyhazi, and Levente Buttyan.
\newblock A survey of interdependent information security games.
\newblock \emph{ACM Computing Surveys (CSUR)}, 47\penalty0 (2):\penalty0 23,
  2014.

\bibitem[Li and Mandayam(2014)]{li2014users}
Tianming Li and Narayan~B Mandayam.
\newblock When users interfere with protocols: Prospect theory in wireless
  networks using random access and data pricing as an example.
\newblock \emph{IEEE Transactions on Wireless Communications}, 13\penalty0
  (4):\penalty0 1888--1907, 2014.

\bibitem[Manshaei et~al.(2013)Manshaei, Zhu, Alpcan, Ba{\c{s}}ar, and
  Hubaux]{manshaei2013game}
Mohammad~Hossein Manshaei, Quanyan Zhu, Tansu Alpcan, Tamer Ba{\c{s}}ar, and
  Jean-Pierre Hubaux.
\newblock Game theory meets network security and privacy.
\newblock \emph{ACM Computing Surveys (CSUR)}, 45\penalty0 (3):\penalty0 25,
  2013.

\bibitem[Miura-Ko et~al.(2008)Miura-Ko, Yolken, Mitchell, and
  Bambos]{miura2008security}
R~Miura-Ko, Benjamin Yolken, John Mitchell, and Nicholas Bambos.
\newblock Security decision-making among interdependent organizations.
\newblock In \emph{21st Computer Security Foundations Symposium}, pages 66--80.
  IEEE, 2008.

\bibitem[Nguyen et~al.(2009)Nguyen, Alpcan, and
  Ba{\c{s}}ar]{nguyen2009stochastic}
Kien~C Nguyen, Tansu Alpcan, and Tamer Ba{\c{s}}ar.
\newblock Stochastic games for security in networks with interdependent nodes.
\newblock In \emph{International Conference on Game Theory for Networks}, pages
  697--703. IEEE, 2009.

\bibitem[Nowzari et~al.(2016)Nowzari, Preciado, and
  Pappas]{nowzari2016analysis}
Cameron Nowzari, Victor~M Preciado, and George~J Pappas.
\newblock Analysis and control of epidemics: A survey of spreading processes on
  complex networks.
\newblock \emph{IEEE Control Systems Magazine}, 36\penalty0 (1):\penalty0
  26--46, 2016.

\bibitem[Ok(2007)]{ok2007real}
Efe~A Ok.
\newblock \emph{Real analysis with economic applications}, volume~10.
\newblock Princeton University Press, 2007.

\bibitem[Osborne and Rubinstein(1994)]{osborne1994course}
Martin~J Osborne and Ariel Rubinstein.
\newblock \emph{A course in game theory}.
\newblock MIT press, 1994.

\bibitem[Pal and Hui(2011)]{pal2011modeling}
Ranjan Pal and Pan Hui.
\newblock Modeling internet security investments: Tackling topological
  information uncertainty.
\newblock In \emph{Decision and Game Theory for Security}, pages 239--257.
  Springer, 2011.

\bibitem[Prelec(1998)]{prelec1998probability}
Drazen Prelec.
\newblock The probability weighting function.
\newblock \emph{Econometrica}, pages 497--527, 1998.

\bibitem[Saad et~al.(2016)Saad, Glass, Mandayam, and Poor]{saad2016toward}
Walid Saad, Arnold~L Glass, Narayan~B Mandayam, and H~Vincent Poor.
\newblock Toward a consumer-centric grid: A behavioral perspective.
\newblock \emph{Proceedings of the IEEE}, 104\penalty0 (4):\penalty0 865--882,
  2016.

\bibitem[Schneier(2008)]{schneier2008psychology}
Bruce Schneier.
\newblock The psychology of security.
\newblock In \emph{Progress in Cryptology--AFRICACRYPT 2008}, pages 50--79.
  Springer, 2008.

\bibitem[Schwartz et~al.(2011)Schwartz, Amin, Gueye, and
  Walrand]{schwartz2011network}
Galina Schwartz, Saurabh Amin, Assane Gueye, and Jean Walrand.
\newblock Network design game with both reliability and security failures.
\newblock In \emph{Allerton Conference on Communication, Control, and
  Computing}, pages 675--681, 2011.

\bibitem[Schwartz et~al.(2013)Schwartz, Shetty, and Walrand]{schwartz2013cyber}
Galina Schwartz, Nikhil Shetty, and Jean Walrand.
\newblock Why cyber-insurance contracts fail to reflect cyber-risks.
\newblock In \emph{Allerton Conference on Communication, Control, and
  Computing}, pages 781--787. IEEE, 2013.

\bibitem[Tversky and Kahneman(1992)]{tversky1992advances}
Amos Tversky and Daniel Kahneman.
\newblock Advances in prospect theory: Cumulative representation of
  uncertainty.
\newblock \emph{Journal of Risk and Uncertainty}, 5\penalty0 (4):\penalty0
  297--323, 1992.

\bibitem[Varian(2004)]{varian2004system}
Hal Varian.
\newblock System reliability and free riding.
\newblock In \emph{Economics of Information Security}, pages 1--15. Springer,
  2004.

\end{thebibliography}
